\documentclass[twocolumn]{aastex62}
\usepackage{natbib}
\usepackage{amsmath, mathtools}
\usepackage{hyperref}
\usepackage{color}
\usepackage{comment}
\usepackage[T1]{fontenc}

\specialcomment{notes}{\begingroup\sffamily\color{red}}{\endgroup}
\includecomment{comment}

\newcommand{\RN}[1]{\textup{\uppercase\expandafter{\romannumeral#1}}}
\newcommand{\sigpos}{\ensuremath{\sigma_{\textrm{pos}}}}
\newcommand{\sigaln}{\ensuremath{\sigma_{\textrm{aln}}}}
\newcommand{\sigadd}{\ensuremath{\sigma_{\textrm{add}}}}

\graphicspath{{./}{figures/}}
\received{Oct 12, 2018}
\revised{Jan 16, 2019}
\accepted{Jan 25, 2019}
\submitjournal{ApJ}

\shorttitle{Galactic Center Astrometry}
\shortauthors{Jia et al.}

\begin{document}

\title{The Galactic Center: Improved Relative Astrometry for Velocities, Accelerations, and Orbits near the Supermassive Black Hole}

\correspondingauthor{Siyao Jia}
\email{siyao\_jia@berkeley.edu}

\author[0000-0001-5341-0765]{Siyao Jia}
\affil{Astronomy Department, University of California, Berkeley, CA 94720, USA}

\author[0000-0001-9611-0009]{Jessica R. Lu}
\affil{Astronomy Department, University of California, Berkeley, CA 94720, USA}

\author[0000-0001-5972-663X]{S .Sakai}
\affil{UCLA Department of Physics and Astronomy, Los Angeles, CA 90095-1547, USA}

\author[0000-0002-2836-117X]{A. K. Gautam}
\affil{UCLA Department of Physics and Astronomy, Los Angeles, CA 90095-1547, USA}

\author{T. Do}
\affil{UCLA Department of Physics and Astronomy, Los Angeles, CA 90095-1547, USA}

\author[0000-0003-2874-1196]{M. W. Hosek Jr.}
\affil{Institute for Astronomy, University of Hawaii, Honolulu, HI 96822, USA}

\author{M. Service}
\affil{Institute for Astronomy, University of Hawaii, Honolulu, HI 96822, USA}

\author[0000-0003-3230-5055]{A.M. Ghez}
\affil{UCLA Department of Physics and Astronomy, Los Angeles, CA 90095-1547, USA}

\author[0000-0002-7452-1496]{E. Gallego-Cano}
\affil{Instituto de Astrof\'{i}sica de Andaluc\'{i}a (CSIC), Glorieta de la Astronom\'{i}a s/n,  18008 Granada, Spain}

\author{R. Sch\"{o}del}
\affil{Instituto de Astrof\'{i}sica de Andaluc\'{i}a (CSIC), Glorieta de la Astronom\'{i}a s/n,  18008 Granada, Spain}

\author[0000-0002-2186-644X]{Aurelien Hees}
\affil{SYRTE, Observatoire de Paris, Universit\'e PSL, CNRS, Sorbonne Universit\'e, LNE, 61 avenue de l$'$Observatoire 75014,  Paris, France}
\affil{UCLA Department of Physics and Astronomy, Los Angeles, CA 90095-1547, USA}

\author[0000-0002-6753-2066]{M.R. Morris}
\affil{UCLA Department of Physics and Astronomy, Los Angeles, CA 90095-1547, USA}

\author{E. Becklin}
\affil{UCLA Department of Physics and Astronomy, Los Angeles, CA 90095-1547, USA}

\author{K. Matthews}
\affil{Astrophysics, California Institute of Technology, MC 249-17, Pasadena, CA 91125, USA}


\begin{abstract}
We present improved relative astrometry for stars within the central half parsec of our 
Galactic Center based on data obtained with the 10 m W.~M.~ Keck Observatory from 1995 to 2017. 
The new methods used to improve the astrometric precision and accuracy include correcting for local astrometric distortions, applying a magnitude dependent additive error, and more carefully removing instances of stellar confusion.
Additionally, we adopt jackknife methods to calculate velocity and acceleration uncertainties. 
The resulting median proper motion uncertainty is 0.05 mas yr$^{-1}$ for our complete sample of 1184 stars in the central 10\arcsec (0.4 pc). 
We have detected 24 accelerating sources, 2.6 times more than the number of previously published accelerating sources,  which extend out to 4\arcsec (0.16 pc) from the black hole. 
Based on S0-2's orbit, our new astrometric analysis has reduced the systematic error of the supermassive black hole (SMBH) by a factor of 2.
The linear drift in our astrometric reference frame is also reduced in the North-South direction by a factor of 4.
We also find the first potential astrometric binary candidate S0-27 in the Galactic center. 
These astrometric improvements provide a foundation for future studies of the origin and dynamics of the young stars around the SMBH, the structure and dynamics of the old nuclear star cluster, the SMBH's properties derived from orbits, and tests of General Relativity (GR) in a strong gravitational field.
\end{abstract}

\section{Introduction}

Located at a distance of 8 kpc, our Galactic Center (GC) has been observed extensively due to its close proximity.
It hosts a compact radio source, Sgr A*, which is associated with a 4$\times10^6 M_{\odot}$ Supermassive Black Hole (SMBH) \citep{Schodel_2002, Ghez_1998, Ghez_2000, Ghez_2005, Ghez_2008, Gillessen_2009}.
A nuclear star cluster (NSC) surrounds Sgr A* and is the only NSC in which individual stars can be resolved with the largest telescopes such as the W.M. Keck Observatory and the Very Large Telescope (VLT) \citep{Ghez_2005, Ghez_2008, Gillessen_2009, Gillessen_2017, Boehle_2016}.
The total mass of the NSC is $\sim$$10^7M_{\odot}$ with a half-light radius $\sim$2--5 pc \citep{Genzel_2010, Schodel_2014, Feldmeier_2017}.
The NSC stars are mostly old, late-type stars ($\sim$1 Gyr).  
However, there is also a population of hot, young stars (4-6 Myr) that  dominates the luminosity in the central parsec with a total mass of $\sim$$10^4M_{\odot}$ \citep{Paumard_2006}. 
So far, spectroscopic observations have identified more than 150 early-type stars, including Wolf-Rayet stars, OB supergiants and OB main-sequence stars \citep{Ghez_2003,Paumard_2006, Bartko_2009, Do_2013,Feldmeier_2015}.

There are still many open issues that observations of the GC, particularly astrometric measurements, can address: 
(1) 
the spectroscopically identified (and thus relatively bright) old stellar population does not show evidence for a cusp, as predicted from theory \citep{Buchholz_2009, Do_2009, Bartko_2010},
(2) the formation mechanism of the young populations is still not well understood \citep{ Levin_2003,Nayakshin_2005,Bartko_2009,Lu_2009}, 
(3) there are many unusual stars that may be the product of tidal interactions with the SMBH or with other stars in the region \citep{Gillessen_2012, Abarca_2014, Witzel_2014}, and
(4) short period stars, such as S0-2, provide an  opportunity to test General Relativity (GR) in a strong field around a SMBH for the first time\citep{Rubilar_2001,Weinberg_2005}.

Late-type populations are presumably dynamically relaxed and are predicted to have a steep core profile, or a cusp \citep{Murphy_1991, Bahcall_1977}.
Contrary to theoretical predictions, some observations have found that the surface density profile of late-type stars in our GC is flat, consistent with the Nuclear Star Cluster having no cusp.
Several plausible dynamical scenarios have been suggested to explain the depletion of late-type giants, such as mass segregation,  stellar collisions, or a recent merger event \citep{Buchholz_2009, Do_2009, Bartko_2010}.
However other studies pointed out that this missing stellar cusp problem is only limited to the brightest few percent stars due to  observational difficulties and the observed density of the faintest stars is actually consistent with the existence of a stellar cusp\citep{GallegoCano_2018, Schodel_2018, Baumgardt_2018}.
Obtaining an unbiased measurement of the stellar distribution will be the key for solving the missing cusp problem.
In particular, significant acceleration detections from astrometric measurements will greatly help with the line-of-sight distance.

The \emph{in-situ} star formation mechanism is now widely accepted, where stars in the vicinity of SMBH are formed from a dense gas disk \citep{Levin_2003, Nayakshin_2005, Paumard_2006, Genzel_2010, Do_2017}.
Detailed measurements of dynamical properties of young stars will help in determining their origin.
Previously publications have found about 20\%  of the young stars rotate in a clockwise disk with an inner edge at 0\farcs8, which is an outcome of the \emph{in-situ} formation mechanism \citep{Levin_2003, Paumard_2006, Bartko_2009, Bartko_2010, Lu_2009, Yelda_2014}. 
This 20\% limit might be a lower limit if we consider binaries \citep{Naoz_2018}.
The slope $\alpha$ of the mass function ($dN/dm \propto m^{-\alpha}$) for this cluster is around 1.7 $\pm$ 0.2 for stars down to 0.5 M$_{\odot}$, much flatter than the traditional Salpeter slope of 2.35 \citep{Lu_2013}.
The dynamical structure of those young stars will greatly help us understand the star formation under extreme environments and even get a direct understanding of the SMBH's properties which cannot be obtained by other methods.

G2 is a dusty red object that was discovered in 2011 \citep{Gillessen_2012}. 
The fact that it is a very red point at L band, but very faint at K band, and shows recombination lines from Br-$\gamma$, makes G2 look like a pure gas \citep{Phifer_2013, Gillessen_2012, Eckart_2013}.
However G2 surprisingly survives its closest approach to Sgr A* in early 2014 \citep{Witzel_2014, Valencia_2015, Abarca_2014, Shcherbakov_2014,Gillessen_2013_G2}, where a pure cloud would get tidally disrupted during periapse passage.
This implies that G2 has to be a compact object, probably a binary merger product.
Better characterization of the binary fraction is needed to answer whether this mechanism is able to explain the star formation in the GC.

S0-2 is one of the most interesting young sources, which is very close to Sgr A* ($<$0.5\arcsec) and has a short period of only 16 years.
In fact, S0-2 has provided a direct evidence for the presence of a SMBH in our GC and  its orbit has been used to calculate the mass and distance to Sgr A* \citep{Ghez_2005, Ghez_2008, Gillessen_2009, Gillessen_2017, Boehle_2016}.
In 2018, S0-2  has reached its second observable closest approach to the SMBH. 
While GR has been thoroughly tested in weak gravitational fields many times \cite{Will_2014, Kramer_2016}, this event marked the first direct test of General Relativity (GR) in a strong gravitational field around a SMBH by measuring relativistic redshift in S0-2's radial velocity \citep{Gravity_2018, Hees_2017}.
Improved astrometry is necessary to characterize the full 3D orbit for S0-2 and will be even more important for future periapse precession tests.

The key for solving those puzzles is accurate and precise astrometric measurements, specifically the positions, proper motions, accelerations and orbits.
Previous observations have been conducted using VLT and Keck telescope: astrometric positions from imaging data reaches an accuracy of $\sim$ 0.3 mas, a factor of 200 smaller than the image resolution in K band \citep{Fritz_2010}.

Achieving precise and accurate astrometry requires cross-matching and transforming stellar positions and photometry into a common coordinate system.
However, this is particularly difficult in the GC region for several reasons \citep{Fritz_2010}.
(1) The GC field is very crowded.  Even with AO, confusion occurs frequently as two or more stars move past each other.
(2) Stars have high proper motions, especially close to the central SMBH, which requires careful consideration when aligning epochs \citep{Yelda_2014,Gillessen_2017}. 
(3) Long time baseline observations are critical for accurate astrometry, but the experimental setup can change with time and the associated distortion needs to be handled carefully \citep{Yelda_2010,Service_2016}. 
In this paper, we implement effective methods to deal with the existing difficulties in accurate relative astrometry measurement, including  confusion, imperfect point spread function (PSF), local distortion of non-standard AO epochs.

The observational data set, including both Speckle and AO, and the data reduction process are presented in Section \ref{sec:obs}.
In Section \ref{sec:align}, the multi-epoch alignment procedure is explained in detail.
With a clean astrometric catalog in hand, we derive proper motions, accelerations and orbital fits in Section \ref{sec:poly}.
The improved astrometry and newly detected accelerating sources are presented in Section \ref{sec:result}. 
We also briefly discuss the potential scientific cases that could benefit from our accurate astrometric measurements in Section \ref{sec:discuss}.
Finally, we summarize our work in Section \ref{summary}.

\section{Observations and Data Reduction}
\label{sec:obs}

\subsection{Observations}

\begin{deluxetable*}{@{\extracolsep{4pt}}llcccccccc}
\tabletypesize{\footnotesize}
\tablecolumns{10} 
\tablewidth{0pt}
\tablecaption{Summary of Speckle Imaging Observations}
\tablehead{
 \multicolumn{2}{c}{Date} & \multicolumn{2}{c}{Frames}	& \colhead{FWHM} & \colhead{Strehl ratio \tablenotemark{b}}	& \colhead{N$_{\textrm{stars}}$}  & \colhead{K$_{\textrm{lim}}$\tablenotemark{c}} & \colhead{\sigpos \tablenotemark{d}} & \colhead{Data Source \tablenotemark{e}} \\
 \cline{1-4}
\colhead{(U.T.)} & \colhead{(Decimal)} \tablenotemark{a} & \colhead{Obtained} & \colhead{Used} & \colhead{(mas)} & \colhead{post-process} & \colhead{} & \colhead{(mag)} & \colhead{(mas)} & \colhead{}
}
\startdata
1995  Jun 9-12  &  1995.439 &  15114 & 5286  &  46  &  0.62  & 380 &  16.4  &  1.47  &  Ref. 1 \\ 
1996  Jun 26-27 &  1996.485 &  9261  & 2336  &  47  &  0.57  & 246 &  15.7  &  3.11  &  Ref. 1 \\ 
1997  May 14    &  1997.367 &  3811  & 3486  &  46  &  0.65  & 358 &  16.4  &  1.29  &  Ref. 1 \\ 
1998  May 14-15 &  1998.366 &  16531 & 7685  &  47  &  0.49  & 251 &  15.9  &  0.92  &  Ref. 2 \\ 
1998  Jul 3-5   &  1998.505 &  9751  & 2053  &  42  &  0.85  & 226 &  16.2  &  0.86  &  Ref. 2 \\ 
1998  Aug 4-6   &  1998.590 &  20375 & 11047 &  46  &  0.65  & 293 &  16.4  &  0.75  &  Ref. 2 \\ 
1998  Oct 9     &  1998.771 &  4776  & 2015  &  47  &  0.52  & 216 &  16.0  &  1.32  &  Ref. 2 \\ 
1999  May 2-4   &  1999.333 &  19512 & 9427  &  45  &  0.78  & 344 &  16.7  &  0.69  &  Ref. 2 \\ 
1999  Jul 24-2  &  1999.559 &  19307 & 5776  &  44  &  0.77  & 303 &  16.8  &  0.37  &  Ref. 2 \\ 
2000  Apr 21    &  2000.305 &  805   & 662   &  48  &  0.46  & 141 &  15.4  &  2.48  &  Ref. 3 \\ 
2000  May 19-20 &  2000.381 &  21492 & 15591 &  45  &  0.62  & 402 &  16.9  &  0.56  &  Ref. 3 \\ 
2000  Jul 19-20 &  2000.548 &  15124 & 10678 &  46  &  0.61  & 410 &  16.7  &  1.10  &  Ref. 3 \\ 
2000  Oct 18    &  2000.797 &  2587  & 2247  &  47  &  0.46  & 209 &  15.8  &  1.70  &  Ref. 3 \\ 
2001  May 7-9   &  2001.351 &  11343 & 6678  &  45  &  0.58  & 344 &  16.3  &  0.95  &  Ref. 3 \\ 
2001  Jul 28-29 &  2001.572 &  15920 & 6654  &  46  &  0.73  & 351 &  16.9  &  0.57  &  Ref. 3 \\ 
2002  Apr 23-24 &  2002.309 &  16130 & 13469 &  46  &  0.65  & 452 &  16.9  &  0.90  &  Ref. 3 \\ 
2002  May 23-24 &  2002.391 &  18338 & 11860 &  44  &  0.74  & 436 &  17.1  &  0.58  &  Ref. 3 \\ 
2002  Jul 19-20 &  2002.547 &  8878  & 4192  &  48  &  0.52  & 300 &  16.5  &  1.23  &  Ref. 3 \\ 
2003  Apr 21-22 &  2003.303 &  14475 & 3715  &  48  &  0.53  & 185 &  15.5  &  1.25  &  Ref. 3 \\ 
2003  Jul 22-23 &  2003.554 &  6948  & 2914  &  46  &  0.65  & 276 &  16.2  &  1.16  &  Ref. 3 \\ 
2003  Sep 7-8   &  2003.682 &  9799  & 6324  &  46  &  0.67  & 356 &  16.6  &  1.80  &  Ref. 3 \\ 
2004  Apr 29-30 &  2004.327 &  20140 & 6212  &  47  &  0.66  & 275 &  16.1  &  0.51  &  Ref. 4 \\ 
2004  Jul 25-26 &  2004.564 &  14440 & 13085 &  47  &  0.61  & 379 &  16.9  &  0.90  &  Ref. 4 \\ 
2004  Aug 29    &  2004.660 &  3040  & 2299  &  49  &  0.79  & 289 &  16.3  &  0.83  &  Ref. 4 \\ 
2005  Apr 24-25 &  2005.312 &  15770 & 9644  &  47  &  0.54  & 282 &  16.3  &  0.70  &  Ref. 5 \\ 
2005  Jul 26-27 &  2005.566 &  14820 & 5642  &  50  &  0.64  & 332 &  16.6  &  1.79  &  Ref. 5 \\ 
\enddata
\label{tab:spe}
\tablenotetext{a}{Decimal year is defined as the Julian Epoch year: 2000.0 + (MJD - 51544.5)/365.25}
\tablenotetext{b}{Strehl ratio reported here is the post-process value from deconvolution method.}
\tablenotetext{c}{K$_{\mathrm{lim}}$ is the magnitude at which the cumulative distribution function of the observed K magnitudes reaches 90\% of the total sample size.}
\tablenotetext{d}{Positional error taken as error on the mean from the three sub-images in each epoch and includes stars with K$<$15.}
\tablenotetext{e}{Data originally reported in (1) \cite{Ghez_1998}, (2) \cite{Ghez_2000}, (3) \cite{Ghez_2005}, (4) \cite{Lu_2005}, and (5) \cite{Rafelski_2007}} 
\end{deluxetable*}


\begin{deluxetable*}{@{\extracolsep{4pt}}llcccccccc}
\tabletypesize{\footnotesize}
\tablecolumns{10} 
\tablewidth{0pt}
\tablecaption{Summary of AO Imaging Observations}
\tablehead{
 \multicolumn{2}{c}{Date} & \multicolumn{2}{c}{Frames}	& \colhead{FWHM} & \colhead{Strehl ratio}	& \colhead{N$_{\textrm{stars}}$}  & \colhead{K$_{\textrm{lim}}$\tablenotemark{a}} & \colhead{\sigpos \tablenotemark{b}} & \colhead{Data Source \tablenotemark{c}} \\
 \cline{1-4}
\colhead{(U.T.)} & \colhead{(Decimal)} & \colhead{Obtained} & \colhead{Used} & \colhead{(mas)} & \colhead{observed} & \colhead{} & \colhead{(mag)} & \colhead{(mas)} & \colhead{}
}
\startdata
2005 Jun 30         & 2005.495   &  10    & 10   &   62 &  0.29   &  794    &   16.2   &     0.39   &  LGSAO; Ref. 8  \\
2005 Jul 31         & 2005.580   &  59    & 31   &   57 &  0.22   &  1753   &   19.0   &     0.25   &  LGSAO; Ref. 7  \\
2006 May 3          & 2006.336   &  127   & 107  &   58 &  0.32   &  1951   &   19.2   &     0.05   &  LGSAO; Ref. 7  \\
2006 Jun 20-21      & 2006.470   &  289   & 156  &   57 &  0.35   &  2438   &   19.5   &     0.08   &  LGSAO; Ref. 7  \\
2006 Jul 17         & 2006.541   &  70    & 64   &   58 &  0.34   &  2165   &   19.3   &     0.09   &  LGSAO; Ref. 7  \\
2007 May 17         & 2007.374   &  101   & 76   &   58 &  0.35   &  2492   &   19.5   &     0.09   &  LGSAO; Ref. 7  \\
2007 Aug 10,12     & 2007.612   &  139   & 78   &   58 &  0.30   &  1877   &   19.1   &     0.08   &  LGSAO; Ref. 7  \\
2008 May 15         & 2008.371   &  138   & 134  &   54 &  0.29   &  2080   &   19.4   &     0.06   &  LGSAO; Ref. 9  \\
2008 Jul 24         & 2008.562   &  179   & 104  &   58 &  0.32   &  2175   &   19.3   &     0.04   &  LGSAO; Ref. 9  \\
2009 May 1,2,4    & 2009.340   &  311   & 149  &   57 &  0.35   &  2297   &   19.4   &     0.04   &  LGSAO; Ref. 9  \\
2009 Jul 24         & 2009.561   &  146   & 75   &   62 &  0.25   &  1699   &   18.9   &     0.09   &  LGSAO; Ref. 9  \\
2009 Sep 9          & 2009.689   &  55    & 43   &   62 &  0.31   &  1920   &   19.1   &     0.11   &  LGSAO; Ref. 9  \\
2010 May 4-5        & 2010.342   &  219   & 158  &   63 &  0.28   &  2027   &   19.2   &     0.06   &  LGSAO; Ref. 9  \\
2010 Jul 6          & 2010.511   &  136   & 117  &   62 &  0.29   &  1950   &   19.1   &     0.08   &  LGSAO; Ref. 9  \\
2010 Aug 15         & 2010.620   &  143   & 127  &   61 &  0.26   &  1819   &   19.1   &     0.07   &  LGSAO; Ref. 9  \\
2011 May 27         & 2011.401   &  164   & 114  &   66 &  0.25   &  1557   &   18.9   &     0.13   &  LGSAO; Ref. 9  \\
2011 Jul 18         & 2011.543   &  212   & 167  &   59 &  0.26   &  2017   &   19.3   &     0.07   &  NGSAO; Ref. 9  \\
2011 Aug 23-24      & 2011.642   &  218   & 196  &   60 &  0.32   &  2354   &   19.5   &     0.05   &  LGSAO; Ref. 9  \\
2012 May 15,18     & 2012.371   &  290   & 201  &   59 &  0.30   &  2256   &   19.4   &     0.05   &  LGSAO; Ref. 10 \\
2012 Jul 24         & 2012.562   &  223   & 162  &   58 &  0.33   &  2317   &   19.5   &     0.06   &  LGSAO; Ref. 11 \\
2013 Apr 26-27      & 2013.318   &  267   & 140  &   68 &  0.22   &  1264   &   18.4   &     0.09   &  LGSAO; Ref. 11 \\
2013 Jul 20         & 2013.550   &  238   & 193  &   58 &  0.33   &  1788   &   19.1   &     0.08   &  LGSAO; Ref. 11 \\ 
2014 May 19         &  2014.380  &  173   & 147  &   64 &  0.28   &  1468   &   18.7   &     0.08   &  LGSAO; Ref. 12 \\
2014 Aug 6          & 2014.596   &  137   & 127  &   56 &  0.33   &  1760   &   19.1   &     0.08   &  LGSAO; Ref. 12 \\
2015 Aug 9-11       & 2015.606   &  288   & 203  &   58 &  0.33   &  1887   &   19.1   &     0.06   &  LGSAO; Ref. 12 \\
2016 May 3          & 2016.338   &  253   & 166  &   60 &  0.31   &  1655   &   18.9   &     0.06   &  LGSAO; Ref. 12 \\
2016 Jul 13         & 2016.532   &  207   & 144  &   60 &  0.26   &  1378   &   18.5   &     0.07   &  LGSAO; Ref. 12 \\
2017 May 4,5       & 2017.343   &  469   & 179  &   59 &  0.31   &  1674   &   19.0   &     0.06   &  LGSAO; Ref. 12 \\
2017 Aug 9-11       & 2017.610   &  213   & 111  &   55 &  0.32   &  1476   &   18.7   &     0.09   &  LGSAO; Ref. 12 \\
2017 Aug 23,24,26 & 2017.647   &  216   & 112  &   61 &  0.29   &  1216   &   18.0   &     0.06   &  LGSAO; Ref. 12 \\
\enddata
\label{tab:ao}
\tablenotetext{a}{K$_{\mathrm{lim}}$ is the magnitude at which the cumulative distribution function of 
    the observed K magnitudes reaches 90\% of the total sample size.}
\tablenotetext{b}{Positional error taken as error on the mean from the three sub-images in each epoch and includes stars 
    with K$<$15.}
\tablenotetext{c}{Data originally reported in (6) \cite{Ghez_2005_ao}, (7) \cite{Ghez_2008}, (8) \cite{Lu_2009}, (9) \cite{Yelda_2014}, (10) \cite{Meyer_2012},  (11) \cite{Boehle_2016}, and (12) this work.} 
\end{deluxetable*}


The central 10$\arcsec$ region of the GC (approximately centered around Sgr A*) has been monitored from the W.~M.~Keck Observatory with diffraction-limited, near-infrared imaging cameras since 1995.
Images used in our multi-epoch astrometric analysis have been obtained in two different modes: 
Speckle imaging from 1995 to 2005 (26 epochs) and laser guide star adaptive optics (LGS-AO) imaging from 2005 to 2017 (30 epochs). 

All Speckle data sets were obtained in the K-band ($\lambda_0$ = 2.2$\micron$) using the Near Infrared Camera (NIRC; \cite{Matthews_1994, Matthews_1996}) on the Keck I telescope with a field of view (FOV) $\sim 5\arcsec \times 5\arcsec$ and a pixel scale of 20 mas. 
Speckle data is taken using very short exposure times (0.1s) to freeze the atmospheric distortion. 
Details of the Speckle observation in Table \ref{tab:spe} can be found in \citet{Ghez_1998, Ghez_2000, Ghez_2005_ao}, \citet{Lu_2005}, \citet{Rafelski_2007} and Z.Chen et al. (in prep). 

Since 2005, we have used the Keck \RN{2} LGS-AO system \citep{vanDam_2006,Wizinowich_2006} with the near infrared camera NIRC2 (PI: K.Matthews) in its narrow-field mode, 
which has a FOV of $\sim 10\arcsec \times 10\arcsec$ and a plate scale of 9.952 mas/pixel \citep{Yelda_2010}. 
After 2014, the adaptive optics system and NIRC2 camera were realigned, and the plate scale changed to 9.971 mas/pixel \citep{Service_2016}.
In this paper, we include new data from 2014 - 2017, increasing the time baseline of the LGS-AO data set by $\sim$30\% (2005 - 2017, compared to 2005-2013 in \cite{Boehle_2016}\footnote{The ``2004 July'' LGS-AO epoch in \cite{Yelda_2014} is dropped in our analysis because of the poor image quality and sensitivity compared with the rest of epochs.}). 
The new LGS-AO data sets (Table \ref{tab:ao}) are obtained in an identical manner and have comparable quality to our previous observations.
Additional details about our observational setup are presented in \citet{Ghez_2005_ao, Ghez_2008, Lu_2009, Yelda_2012, Yelda_2014} and \cite{Boehle_2016}.

\subsection{Image Processing}
\label{sec:starlist}

\cite{Boehle_2016} combined the individual Speckle frames using a ``holography'' method \citep{Schodel_2013} rather than the traditional shift-and-add method \citep{Ghez_2003}. 
We improved this algorithm by using multiple reference stars,  removing nearby confusion sources and subtracting sky background when extracting PSF.
To estimate the positional errors, 100 realizations of each epoch were created using a bootstrap sampling with replacement to combine frames for that epoch. 
\textit{Starfinder} is run on each of these realizations and the standard deviation of the positional measurements of stars are adopted as the uncertainty. 
We find that this method provides a more robust estimate of astrometric and photometric uncertainties (Z.Chen et al, in prep). 
The  new holography method allows us to use more exposures and increases the sensitivity and field of view of the Speckle images, resulting in more stars detected at fainter magnitude.
On average, 309 stars are detected in Speckle data sets down to a 90\% limiting magnitude of K$_{\rm lim}$ = 16.4, with average position error of 1.1 mas in each direction. 

The NIRC2 images were reduced using our standard NIRC2 reduction pipeline, which includes corrections for geometric distortion and differential atmospheric refraction. 
We used a new photometric calibration described in \cite{Gautam_2019} to recalibrate all LGS-AO data sets. 
For observations between 2004-2013, we use the distortion solution from \cite{Yelda_2010}, while for observations obtained in 2014 and later, we use the new static distortion map from \citet{Service_2016}. This change in optical distortion is a result of changes in the alignment of the AO system \citep{Service_2016}. 
The output of the NIRC2 pipeline is a single combined image for each epoch of data along with three sub-set images used for error analysis \citep{Ghez_2005}.
Positional uncertainties, $\sigma_{\mathrm{pos}}$, is the error on mean of the positions in three subset images for each star. 
Starlists containing astrometry and photometry were extracted using a PSF fitting algorithm \emph{StarFinder} \citep{Diolaiti_2000, Yelda_2014}. 
On average, 1850 stars are detected in AO data sets down to a 90\% limiting magnitude of K$_{\rm lim}$ = 19.0, with average position error of 0.09 mas in each direction. 
This exquisite precision is a result of the high signal-to-noise ratio (SNR). 
For a star at K=15.5 mag, the typical SNR is approximately 3000 and 8000 for Speckle and AO separately. 
Given that the FWHM=45-65 mas and the lowest possible centroiding error is $\sigma_{pos} \sim FWHM/SNR$, we could potentially reach positional uncertainties as low as 0.01 mas.
However, the astrometric precision and accuracy is mostly limited by systematic errors due to uncertainties in the point-spread function (PSF) and noise from the halos of the surrounding stars \citep{Trippe_2010}.

\section{Multi-Epoch Astrometric Alignment}
\label{sec:align}

The starlists from each epoch must be aligned (i.e. transformed and cross-matched) into an \emph{absolute reference frame} in order to measure accurate proper motions, accelerations, and orbits.
We define an \emph{absolute reference frame} as described in \citep{Sakai_2019} by measuring the proper motions of a set of IR stars that have been accurately transformed into a radio Sgr A*-rest frame. 
The resulting catalog of \emph{astrometric secondary standards} contains $\sim$800 bright stars evenly distributed over the central 20$\arcsec$ $\times$ 20$\arcsec$. We adopt this catalog of positions and velocities as our \emph{absolute reference frame}, and all of the deep, high-precision observations of the central 10$\arcsec$$\times$10$\arcsec$ will be transformed into this common frame.

Our starlists are transformed in two steps: 

1) We choose a reference epoch (2009 May 04, one of the well measured AO epochs), and transform starlists from other epochs into this reference epoch coordinate system by fitting a second order bivariate polynomial transformation. 
In this step, the set of \emph{reference stars} used to calculate the transformation is selected from the catalog of \emph{astrometric secondary standards} (see \S\ref{sec:ref_ao} and \S\ref{sec:spe_align}). The \emph{reference stars} are first propagated from the reference epoch to other epochs using their known velocities from the \emph{absolute reference frame} and the position angle and plate scale of the reference epoch.
Then stars are matched between the propagated reference epoch and the starlist using a matching radius of 40 mas (i.e. $\sim$4 pixel for AO and $\sim$2 pixel for Speckle, see \cite{Lu_2009} and \cite{Yelda_2010} for more details on matching).
The best-fit transformation is then calculated by minimizing the residuals between the predicted positions and measured positions at each epoch for \emph{reference stars}.
At this point, all epochs are still in units of pixel in the reference-epoch coordinate system.

2) We then align the reference epoch coordinate (2009 May 04) to the \emph{absolute reference frame} in order to transform all pixel positions into arcsec on the sky. 
This transformation is also accomplished by a second-order polynomial transformation in an iterative process using the 50 brightest stars first, and then increasing the number of stars in successive iterations as more stars are matched with a better transformation. 
This iterative process stops when the number of matched stars do not increase significantly.

We have implemented a number of improvements  to our multi-epoch alignment process.
First, in order to assess the impact of instrument and AO changes in the 24 years of astrometry, we establish a baseline alignment using a subset of data with identical instrumental and AO setups. This alignment of data from 2006 to 2014, which we call the \emph{06-14} alignment, represents the highest precision subset of the astrometry. We use this data set to characterize systematic errors and stellar confusion events as described below.

In \S\ref{sec:06_align}, we add a magnitude-dependent additive error to increase the accuracy of the astrometric errors; and we develop an algorithm to detect artifact sources on the edge of AO images due to elongated PSFs.
In \S\ref{sec:AO_align}, we incorporate the remaining AO data by deriving local distortion maps to compensate for different experimental setups (e.g.~NIRC2 data after 2014).
Then, Speckle data is added to the alignment in \S\ref{sec:spe_align}.
Finally, we utilize new thresholds to remove instances of stellar confusion in \S\ref{sec:matching}. 

\subsection{06-14 Alignment}
\label{sec:06_align}

\subsubsection{Magnitude Dependent Additive Errors} 
\label{sec:additive error}

The uncertainty of each star's position is comprised of 3 components: 
(1) \sigpos: the measurement precision on each star's position, 
(2) \sigaln: the uncertainty in the transformation process for each star, 
(3) \sigadd: an additive error term to capture additional sources of error. 
\sigpos\ is calculated as the error on the mean of a star's position measured from the three subset images as described in \S\ref{sec:starlist}.
The alignment error, \sigaln\, is estimated from a half-sample bootstrap analysis that repeats the alignment process 100 times with random sets of reference stars and is taken as the standard derivation of each star's position over the bootstrap samples. 
We found that the combination of \sigpos\ and \sigaln\ alone yields a $\chi^2$ distribution on the acceleration fits  with an unexpected tail of high values 
that are inconsistent with the standard $\chi^2$ distribution. 
This is rectified with \sigadd.  

Here, $\chi^2$ is defined as: $\chi_{\vec{p}}^2 = \sum\nolimits_{i \in epochs} (\vec{p}_i - \vec{p}_{mod,i})^2/ \vec{\sigma}_{p,i}^2$, where $\vec{p}_i$ is the position at epoch i, $\vec{p}_{mod,i}$ is the linear/acceleration fit at epoch i and $\vec{\sigma}_{p,i}$ is the positional uncertainty at epoch i (the quadratically-summed combination of \sigpos, \sigaln\; and \sigadd). 
The linear/acceleration fit is discussed in detail in \S\ref{sec:poly}.
We will use $\chi_{acc}^2$ for $\chi^2$ of acceleration fits and $\chi_{lin}^2$ for $\chi^2$ of linear fits.
The fit in the x and y directions are independent of each other, so there are $\chi^2$ values in each direction.
Ultimately, the $\chi^2$ distribution of a sample of stars is an important factor for determining the quality of our analysis.

The additive error term, \sigadd, was previously determined to be 0.10 mas for AO data \citep{Yelda_2014} and was assumed to be constant with time, position, and brightness. 
However, the most likely source of additional astrometric error is from inaccurate estimation of the PSF wings, which predominantly impacts the astrometry of neighboring sources. 
The impact of this effect is largest when the brightness of the neighboring source becomes comparable to the flux in the wings of the many surrounding bright stars.
We improved the determination of \sigadd\; by implementing a magnitude-dependent error term for AO data that is added in quadrature, as described below \cite[see also][]{Fritz_2010, Clarkson_2012}.

\paragraph{Defining the Sample of Good Stars}
In order to evaluate the $\chi^2_{acc}$ distribution and determine the optimal additive error, \sigadd, we define a sample of \emph{good stars} chosen as those that 
(1) were detected in all 22 epochs of the \emph{06-14} alignment,
(2) showed no source confusion in any epoch, as defined in \S \ref{sec:matching}, 
(3) were located between 0\farcs8 and 10$\arcsec$ from the SMBH in order to eliminate stars close to the SMBH that show high-order motion beyond acceleration, 
(4) were not outliers with large acceleration uncertainties, and
(5) had $\chi^2_{\textrm{acc}}$ smaller than 95 (5 times the \emph{dof}), where \emph{dof} is defined as the number of detections minus the number of free parameters.
 For example, here we have 22 epochs in total, and we use an acceleration fit with 3 free parameters per axis thus the \emph{dof} per axis equals to 22-3=19.
The resulting sample of 370 \emph{good stars} will be used to  evaluate the $\chi^2$ distribution for different additive errors. 

\begin{figure}[htp]
  \plotone{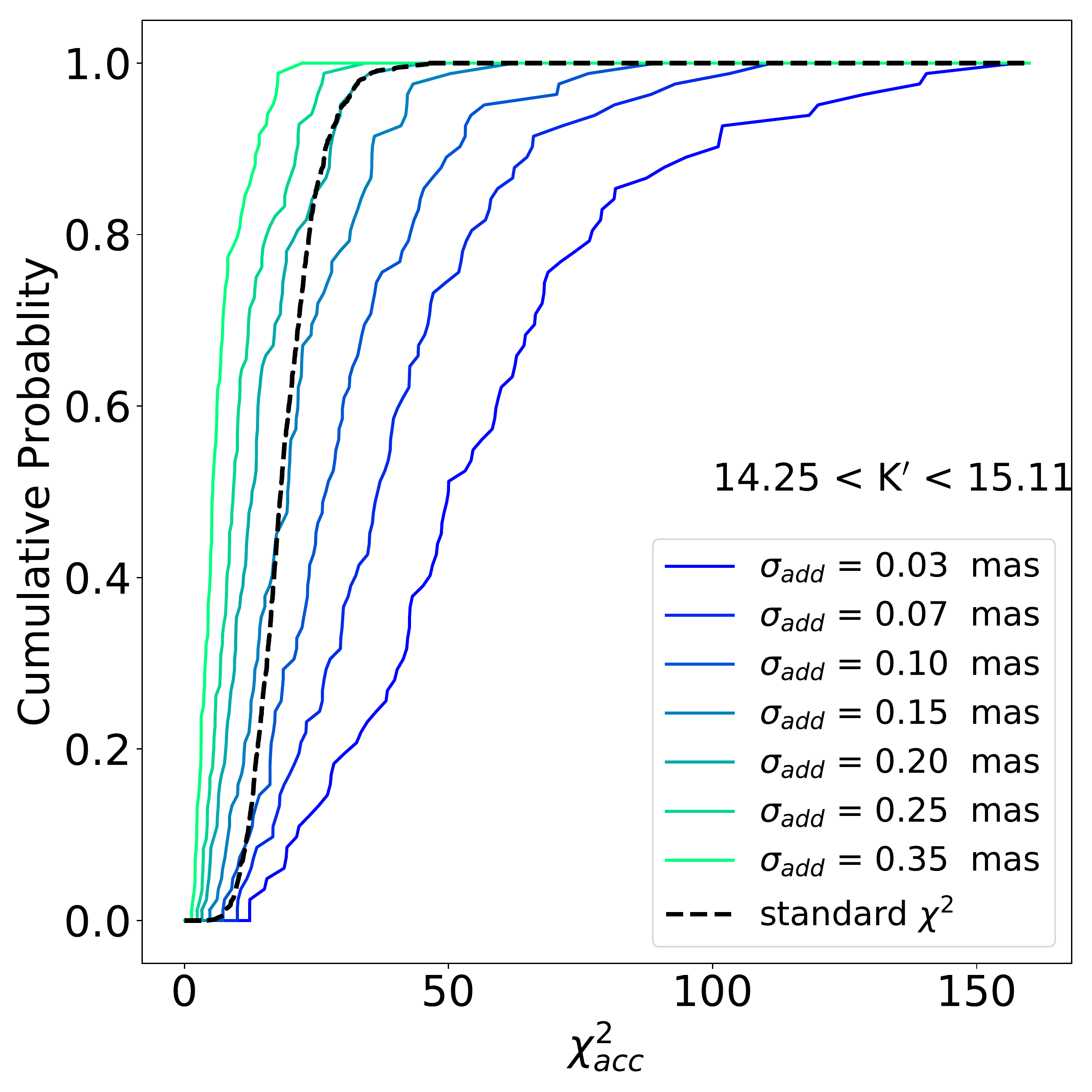}
  \caption{The cumulative $\chi^2_{acc}$ distribution for different \sigadd\; from the \emph{good stars} in the magnitude bin, $14.25 < K < 15.11$. 
   Different colored lines represent different \sigadd\; added to the positional uncertainty in quadrature. 
   With larger \sigadd, the $\chi^2_{acc}$ distribution shifts to smaller values.
  The standard cumulative $\chi^2_{acc}$ distribution with \emph{dof}$=19$ is plotted in the black dashed line.
  The optimal \sigadd\; of 0.15 mas produces a $\chi^2_{acc}$ distribution that most closely matches the expectation. }
  \label{fig:chi2_change}
\end{figure}

\begin{figure}[htp]
  \plotone{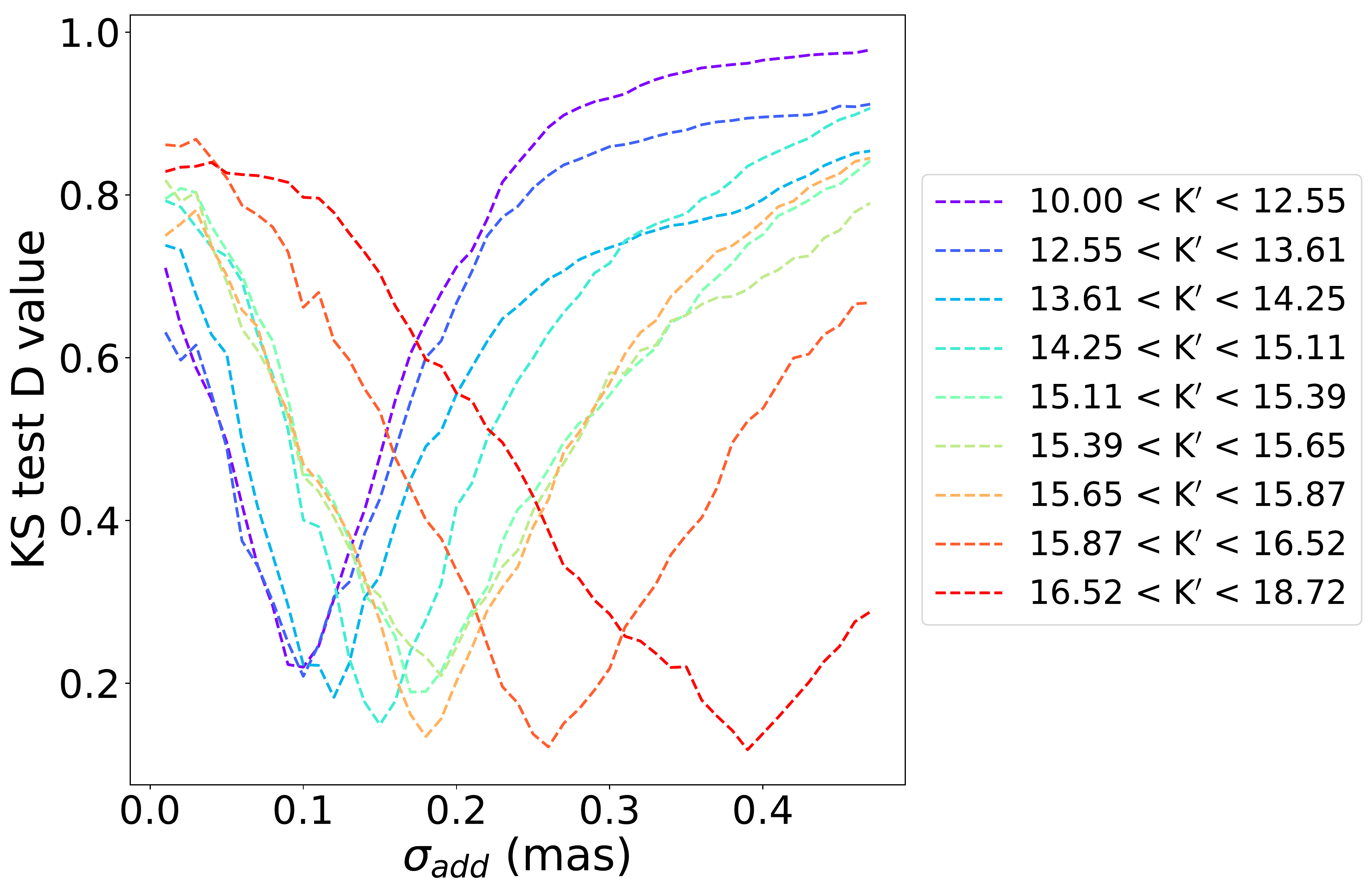}
  \caption{ The KS-test between the observed $\chi^2_{acc}$ distribution and a standard $\chi^2_{acc}$ distribution. 
           The KS test D value is plotted as a function of additive error. 
           The smaller the D value is, the better agreement between the observed and standard $\chi^2_{acc}$ distribution.
           We can see that for fainter stars, they need larger additive error to fit the standard $\chi^2_{acc}$ distribution.}
	\label{fig:chi2_ks}
\end{figure}

\paragraph{Calculating the Additive Error}
To find the optimal \sigadd\; term, we perform an \emph{06-14} cross-epoch alignment with additive errors from 0.01 mas to 0.47 mas in steps of 0.01 mas. 
We then divide the \emph{good stars} into 9 magnitude bins with $\sim$41 stars per bin. 
The magnitude boundaries are: 10, 12.55, 13.61, 14.25, 15.11, 15.39, 15.65, 15.87, 16.52, and 18.72.  
For each magnitude bin and each trial additive error, the $\chi_{acc}^2$ distribution of \emph{good stars} is calculated (Figure \ref{fig:chi2_change}). 
Increasing the additive error causes the $\chi_{acc}^2$ distribution to shift to smaller values. 

The optimal additive error should exhibit a $\chi_{acc}^2$ distribution that closely matches the standard $\chi_{acc}^2$ distribution with dof=19.
We perform a two-sample Kolmogorov-Smirnov test (KS-test) between the observed $\chi_{acc}^2$ distribution and a standard $\chi_{acc}^2$ distribution for each magnitude bin \citep{KS_test}.
The KS test statistic (D value), which quantifies a distance between the observed $\chi_{acc}^2$ distribution  and the cumulative distribution function of the standard $\chi_{acc}^2$ distribution, is plotted as a function of additive error in Figure \ref{fig:chi2_ks}. 
The smaller KS statistic D value suggests a better agreement with the theoretical predicted $\chi_{acc}^2$ distribution.
There is a clear trend showing that fainter stars require larger additive errors. 

\begin{figure}[htp]
  \plotone{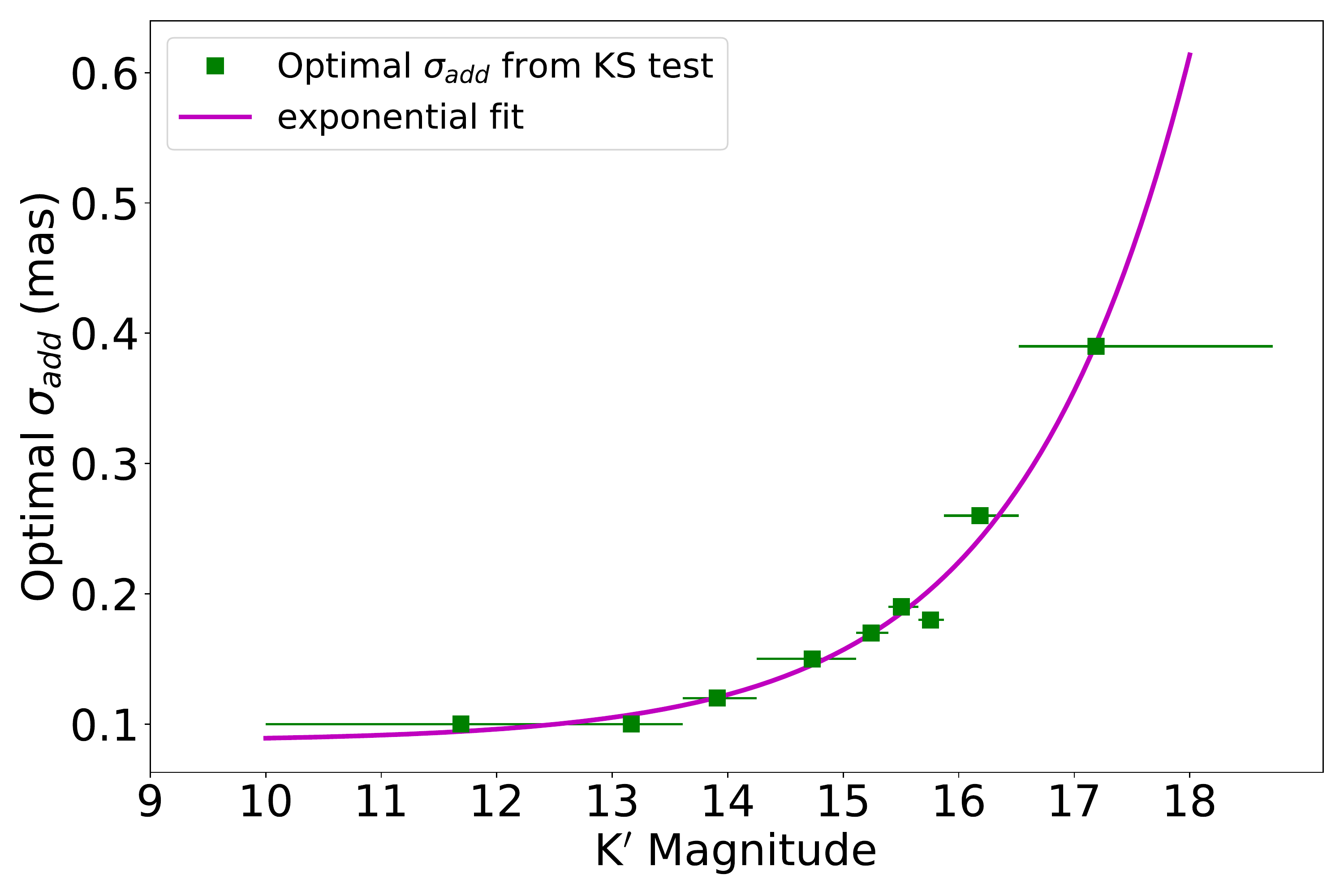}
  \caption{The optimal astrometric additive error, \sigadd, as a function of brightness. 
           The green squares are the optimal \sigadd  \ from minimizing KS test D value from Figure \ref{fig:chi2_ks} .
           The error bar shows the range of each magnitude bin, and the mean of each magnitude bin is showed
           on the green square.
           Based on the trend, we use an exponential function shown in Equation \ref{equ:additive_error} to fit the 
           observed \sigadd \ as a function of K$'$.
           The best-fit exponential relation is shown in the magenta line.}
  \label{fig:bestfit_add}
\end{figure}

In Figure \ref{fig:bestfit_add}, we plot the best fit additive error in each magnitude bin. 
We derive an analytic function for the additive error as a function of magnitude given as
\begin{equation}
  \sigma_\textrm{add} (mas) = 8\times 10^{-5} \times e^{0.7 \times (K' - 4.9)} + 0.1
  \label{equ:additive_error}
\end{equation}

\begin{figure}[htp]
  \plotone{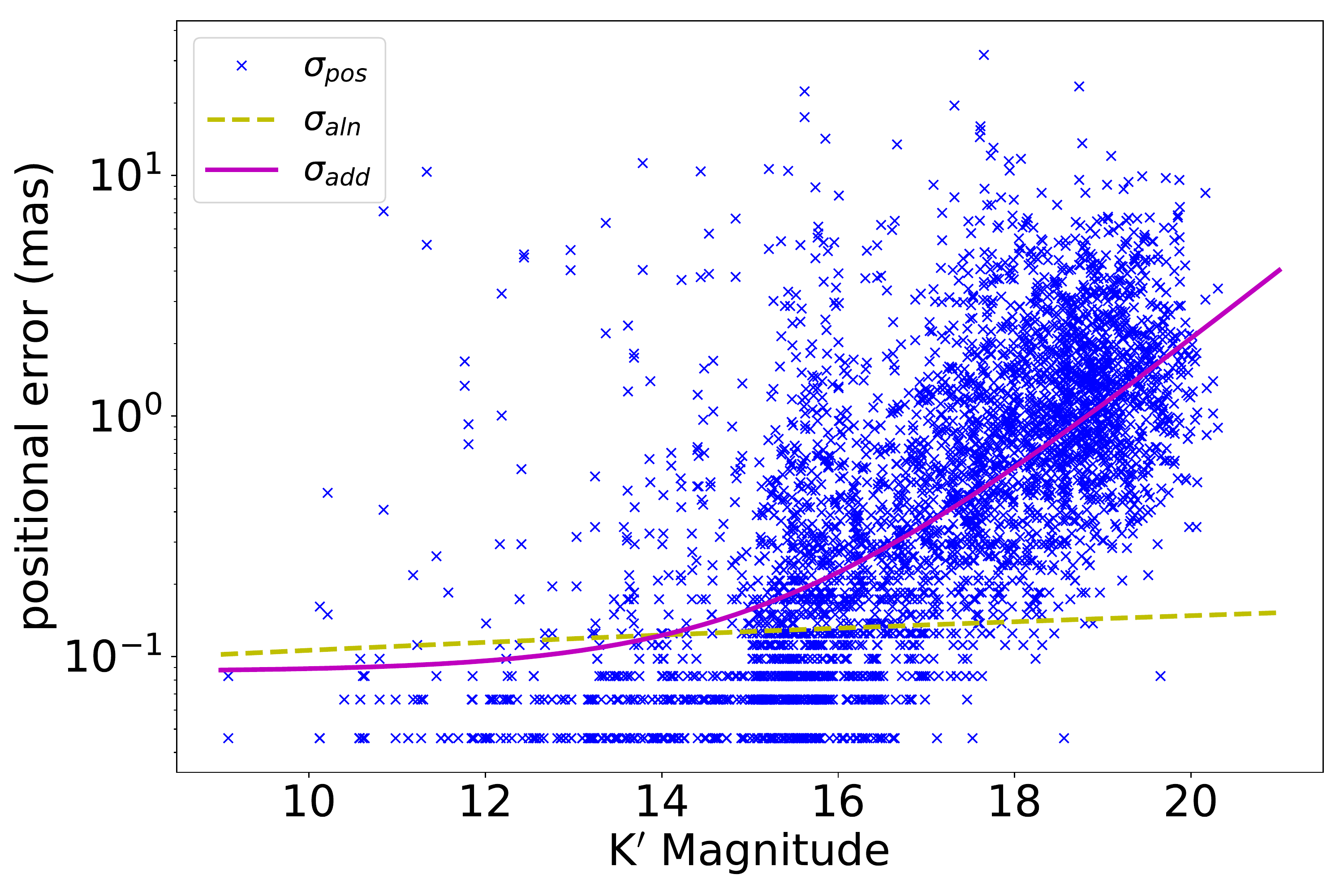}
  \caption{Different sources of positional uncertainty as a function of K$'$ magnitude for 2006 May 03.
           The blue crosses represents \sigpos \  in both the x and y direction. 
           The magenta solid line represents \sigadd, which is the same as showed in Figure \ref{fig:bestfit_add}.
           The yellow dashed line represents \sigaln, which comes from the alignment transformation uncertainty.
           The \sigaln \ curve stays around 0.1 mas for all stars, almost independent of K$'$.
           \sigpos \ and \sigadd \ is 0.1 mas for the bright stars and then increases rapidly for fainter stars.
           The turnover happens around K$'$ = 15.
           Note that our positional uncertainty measurement precision is 0.01 mas, so the smallest
           \sigpos \ looks discrete.}
  \label{fig:err_mag}
\end{figure}

With this magnitude-dependent \sigadd, we now evaluate the contribution of each source of positional uncertainty to the total.
The 2006 May 03 epoch (one of the \emph{06-14} epochs) is used as an example to show how \sigpos, \sigaln, and \sigadd\ change with magnitude in Figure \ref{fig:err_mag}.
We can see that both \sigpos  \ and \sigadd \  increase with magnitude, almost exponentially, and they are comparable with each other. 
In contrast, \sigaln \ captures the transformation uncertainty and is around 0.1 mas with a weak dependence on magnitude. This is likely due to the fact that \sigaln\; is slightly larger in the outskirts of the field, where the faint stars are more easily detected away from the central concentration of bright stars.

\begin{figure}[htp]
\plotone{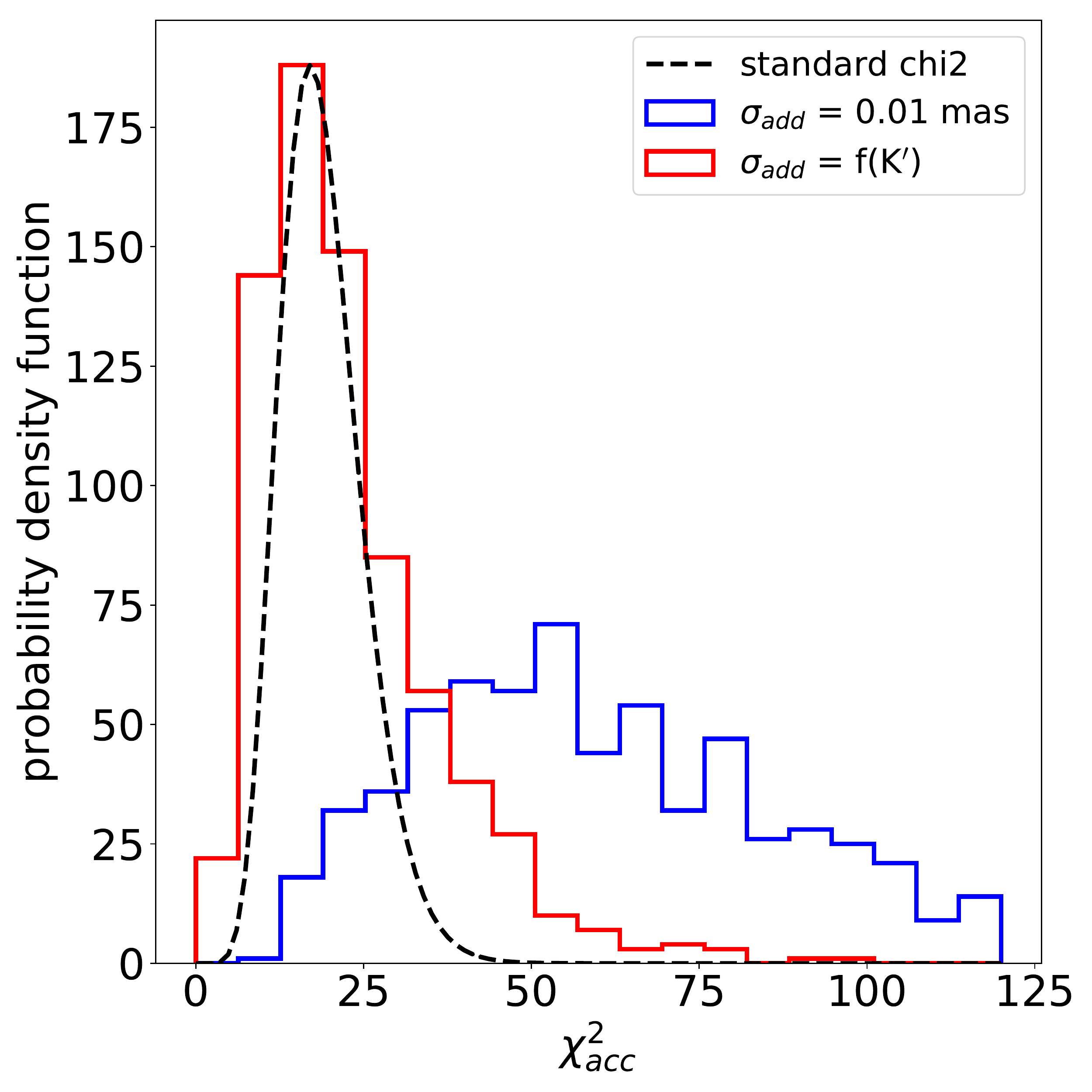}
\caption{The $\chi_{acc}^2$ probability density function (PDF) for the \emph{06-14} alignment.
The blue histogram is the original $\chi_{acc}^2$ distribution with 0.01 mas additive error,  which is approximately the $\chi_{acc}^2$ distribution without any additive error.
The red histogram is the final $\chi_{acc}^2$ distribution with optimal magnitude dependent \sigadd\  from Equation \ref{equ:additive_error}. 
The black dash line is the standard $\chi_{acc}^2$ distribution with \emph{dof} = 19, normalized to the maximum value of observed $\chi_{acc}^2$ PDF.}
\label{fig:chi2}
\end{figure}

This magnitude-dependent additive error is incorporated in for all other AO epochs.
The $\chi_{acc}^2$ distribution for \emph{06-14} alignment with this magnitude dependent additive error is plotted in Figure \ref{fig:chi2} and shows good agreement with the standard $\chi_{acc}^2$ distribution.

\subsubsection{Reference Stars for AO}
\label{sec:ref_ao}
The selection of \emph{reference stars} are critical as they are used to bring each epoch of observations to the Sgr A*-radio rest frame of reference.
Ideally, we would like to maximize the number of  reference stars to yield more accurate transformations between individual epoch and reference epoch (2009 May 04).
However, in order to define the positions of these stars for each epoch, we want the reference stars to move linearly with accurate velocity measurements, since we assume a linear motion model to propagate between epochs.

From \S\ref{sec:additive error}, the sample of \emph{good stars} consists of 370 stars and their $\chi^2_{acc}$ distribution is well-behaved. 
Among these 370 \emph{good stars}, 18  have non-SMBH accelerations:  significant tangential accelerations, significant positive radial accelerations, or negative radial accelerations that are higher than physically allowed from the SMBH (see details in \S\ref{sec:acc}).
These non-SMBH accelerations likely come from confusion or binarity, so we exclude these non-SMBH-acceleration sources, leaving a clean sample of 352 stars. 
In order to obtain accurate velocity measurements in an absolute reference frame, we find the intersection of these 352 \emph{good stars} and the \emph{astrometric secondary standards}, which gives us 141 stars.
These 141 stars are used as \emph{reference stars} in order to derive the coordination transformations for all AO epochs.

\subsubsection{Artifact Sources on the Edge of AO Epochs}
\label{sec:fake}

\begin{figure}[htp]
\plotone{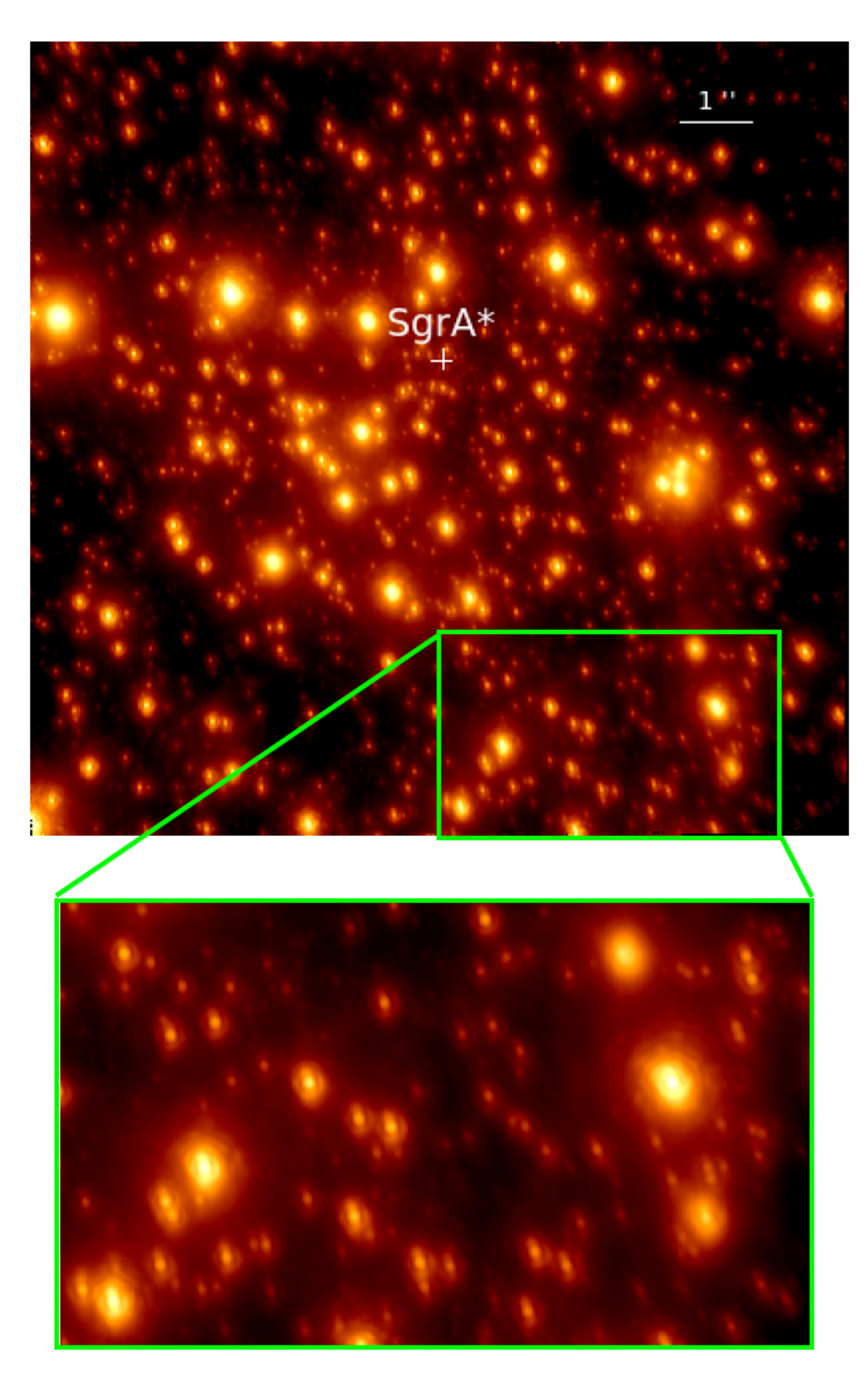}
\caption{An image for epoch 2011 July. 
Stars in the central region are more circular as compared to stars in the lower right corner, which are all elongated in a similar direction.
The PSF is most elongated in the direction away from the tip-tilt star.}
 \label{fig:psf_edge}
\end{figure}

The Keck AO observations deliver near-diffraction-limited spatial resolution; however, the PSF for the AO images is highly structured and varies across the field of view. 
We found that, near the edges of images from AO epochs, the PSF becomes elongated and structures in the first Airy ring are sometimes identified as a separate source.  
Figure \ref{fig:psf_edge} shows an example.

\begin{figure}[htp]
\plotone{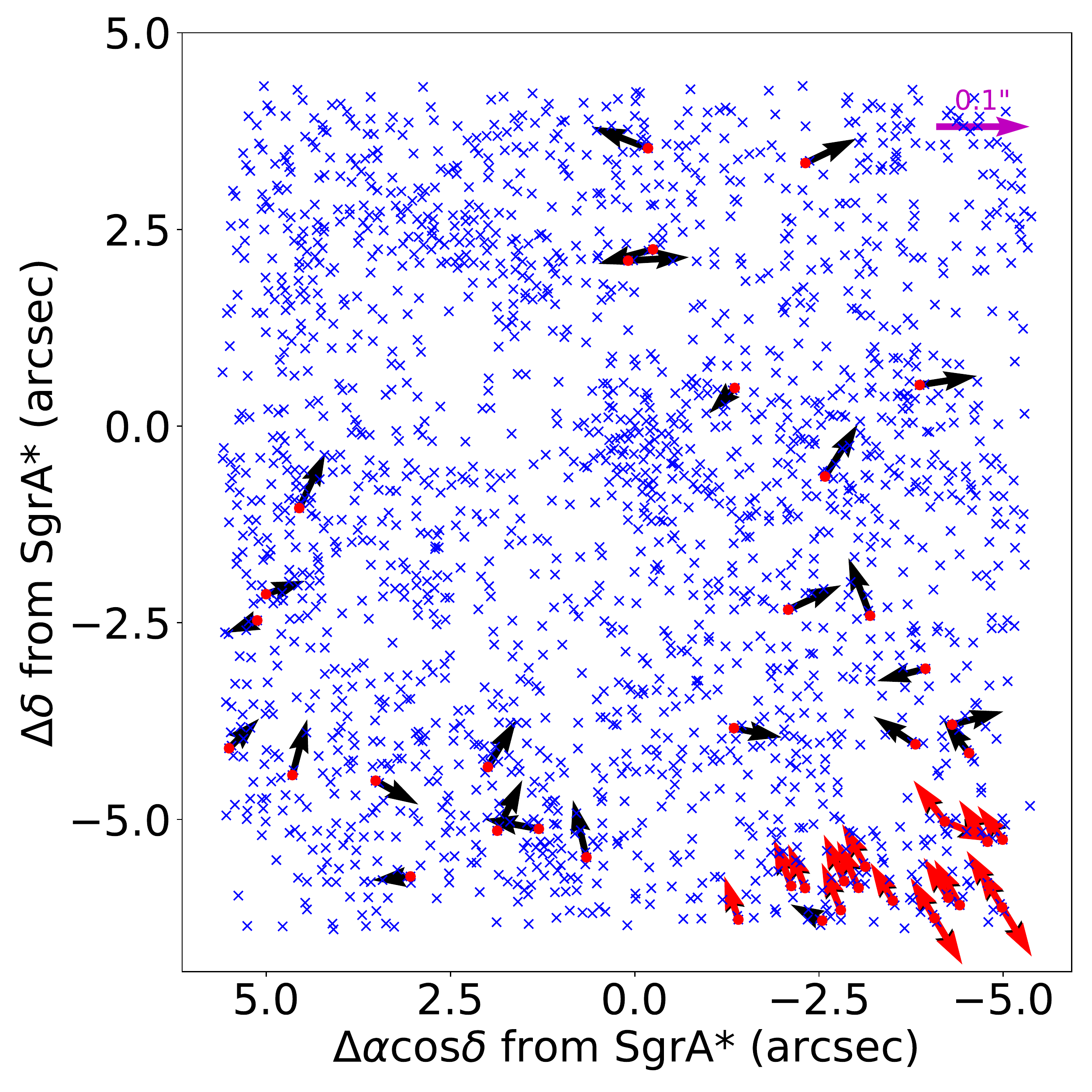}
\caption{Artifact sources due to PSF field variability.
Each blue cross is a star detected in epoch 2011 July 18.
Then each star is paired up with every other star and the red points are marked when a pair of star has a positional offset smaller than 70 mas and a proper motion difference smaller than 3 mas yr$^{-1}$.
The position offsets between those pairs are plotted in the black and red arrows.
If a pair is just a coincidence, then the arrows should be randomly distributed; however, we see many arrows in the lower-right corner have the same direction, which suggests that PSF elongation has created these artificial sources.
Those arrows with the same direction are colored in red as artifact sources which are excluded from the final sample.}
\label{fig:fake}
\end{figure}

Active work is underway to account for PSF variations across the field \citep[e.g. the AIROPA project][]{Witzel_2016}.
However, in this paper, we use a simple yet straightforward way to mark and remove those artifact sources. 
Artifact sources have some common properties: 
(1) artifact sources are usually within 70 mas of their primary sources. 
(2) because the artifact sources are typically in the same relative position compared to their primary sources, the proper motions for the artifact sources and primary sources are similar. 
(3) the PSF elongation is usually in the same direction as the separation vector between the primary and artifact sources.
Combined, these three properties enable us to find potential artifact sources coming from PSF field variability. 
Figure \ref{fig:fake} uses epoch 2011 July 18 as an example to show how this correction works.
First, we need to find pairs of sources based on the first 2 properties: 
we select every pair of stars that are within 70 mas of each other and have similar proper motions ($\delta \mu<$ 3 mas yr$^{-1}$). 
Each blue cross is a detected star in epoch 2011 July 18, and the red points are the pairs pass our selection criteria.
The positional offset between each pair is shown as black and red arrows.
The offsets are all in similar directions in the lower right corner, 
which is a strong evidence that they are artifact sources coming from PSF wings, and this
agrees with what we see in Figure \ref{fig:psf_edge}.
For those stars that have the same offset direction, we will mark them as PSF artifact sources,
which is shown in the red arrows in Figure \ref{fig:fake}.
Among 2729 sources, 88 sources are found to be artifact sources in 26 AO epochs, accounting for about 3.2\% of the total number of sources. 
The artifact sources are excluded from our final sample.

\subsubsection{Update Matching Velocity from \emph{06-14} Alignment}
\label{sec:update_v}

The epochs from 2006 to 2014 are taken with exactly the same instrumental setup.
As a result, these data give a much more precise velocity measurements compared to \cite{Yelda_2014}.
Therefore, we use the velocity from \textit{06-14} alignment to update the velocities for those stars that are not \emph{astrometric secondary standards} but still need to be matched based on their matching velocity.
Matching velocities are updated for stars if:
1) Their projected distance to Sgr A* is larger than 0\farcs4. This is because stars within 0\farcs4 need a model beyond simple linear motion.
2) Their K band magnitude is brighter than 16 magnitude.
3) Their velocity uncertainty from \textit{06-14} alignment is smaller than 2 mas yr$^{-1}$.
556 stars' velocities are updated from the \textit{06-14} alignment.
Comparing with \citet{Yelda_2014}, the median velocity uncertainty for these 556 stars is reduced from 0.14 mas yr$^{-1}$ to 0.03 mas yr$^{-1}$, by almost a factor of 5.

\subsection{Adding Other AO Epochs: Local Distortion}
\label{sec:AO_align}

Among the 30 AO epochs, 8 were not taken in the standard \textit{06-14} setup including the 2005, 2015, 2016, and 2017 epochs. 
They have higher order residual distortion from changes in the AO system optics, so the standard second order polynomial transformation is insufficient to place these images into a common reference frame. 
Therefore, we need to make local distortion maps for these epochs.
Our local distortion maps are calculated based on residuals, which is different from geometric distortions based on the on-sky measurements in \cite{Yelda_2010} and \cite{Service_2016}.
The way we calculate the distortion map is described as follows.

\begin{figure*}
\centering
\includegraphics[width=\textwidth]{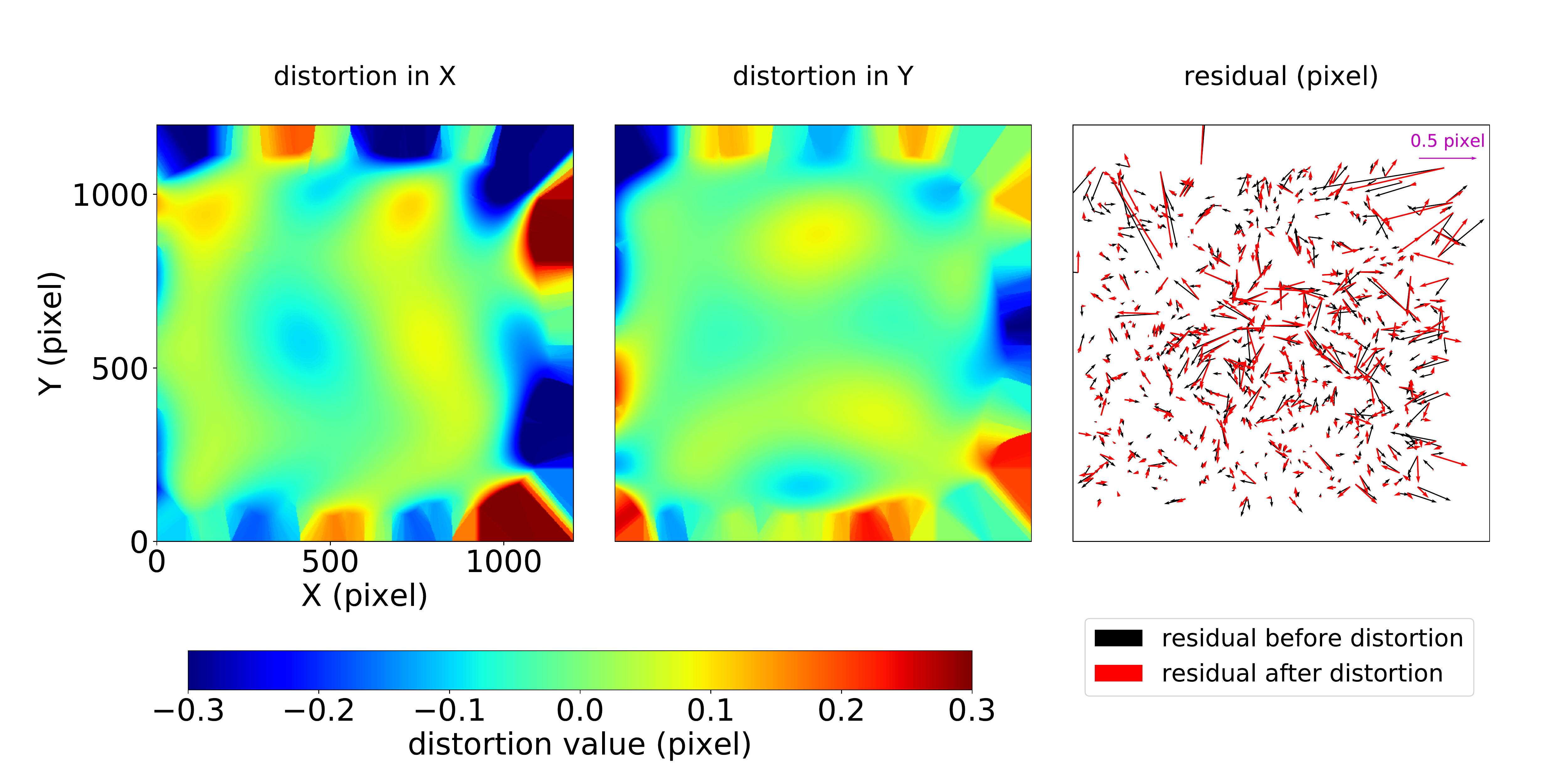}
\caption{The local distortion map for 2016 May 03 in the $x$ direction ({\em left}) and $y$ direction ({\em middle}), color coded by the distortion value.  
On the {\em right}, the individual star's residuals from the multi-epoch acceleration fit are shown both before ({\em black}) and after ({\em red}) the local distortion solution is applied.
The arrows are the offset between the \textit{06-14} alignment propagated position and the 2016 May 03 position.
The distortion map is the Legendre transformation calculated from the black arrows. 
Note that there are very few stars on the edge, so we use the nearest distortion value  to extrapolate into those regions. 
The red arrows are reduced by a factor of 1/3 compared to the black arrows, especially at corners, indicating the effectiveness of the distortion map.}
\label{fig:distmap}
\end{figure*}

First, the \textit{06-14} alignment is used to fit acceleration models for stars within 0\farcs8 and 
linear models for stars outside of 0\farcs8 from Sgr A*. With the acceleration/linear fit, 
all stars are propagated to each of the non-standard epochs listed above. 
The differences between the propagated positions and the observed positions at that epoch are used to estimate the local distortion map. 
When calculating the distortion map, the following cuts are made to reduce noise in the final distortion map: 
1) Only stars brighter than 17 mag are used. 
2) Outliers are removed by dividing the detected stars into 6 $\times$ 6 spatial around the position of Sgr A*, calculating the mean and standard deviation of the distortion in each box, and trimming stars with offsets that are more than 2.5$\sigma$ from the mean.
This gives us the final sample that is used to calculate the local distortion map.
On average, 530 stars are used in the final sample.

A high order Legendre transformation is fit to the residuals to construct the local distortion map.
To calculate the uncertainty of the local distortion map, we use the standard deviation among 100 local distortion maps estimated from a full-sample bootstrap with replacement. 
The distortion and its uncertainty can become very large on the edge of the distortion map as this region is sometimes outside the field of view for a given epoch and the distortion must be extrapolated from very few stars.
Therefore, for those ``edge'' values, whose distortion or distortion uncertainty is larger than 0.3 pixel, we use the nearest ``non-edge'' value to replace it.

Finally, we need to find the best Legendre transformation order for our distortion map. 
The residuals will always improve as we increase the transformation order; but the number of free parameters also increases for a higher order transformation. 
So we use the F-ratio test to find when an increase in the Legendre transformation order no longer {\em significantly} improves the residuals.
This is done by finding the point when (1-p) value for F-ratio approaches 0. 
The F-ratio is calculated when increasing Legendre transformation order, and the p-value is the probability of obtaining this F-ratio (see Equation (B2) in \cite{Lu_2016}). 
Lower p-values indicate more significant benefits to increasing the order of the transformation polynomials.
A similar test has been done in \cite{Lu_2016}.

\startlongtable
\begin{deluxetable}{lcc}
\tabletypesize{\footnotesize}
\tablewidth{0pt}
\tablecaption{Local Distortion Summary \label{tab:local_dist}}
\tablehead{
\colhead{Date} & \colhead{Distortion in X} & \colhead{Distortion in Y}\\
\colhead{} & \colhead{(pixel)} & \colhead{(pixel)} 
}
\startdata
2005 Jun 30  & 0.028 $\pm$ 0.025 & 0.050 $\pm$ 0.039 \\
2005 Jul  31 & 0.029 $\pm$ 0.016 & 0.051 $\pm$ 0.021 \\
2015 Aug  10 & 0.033 $\pm$ 0.017 & 0.030 $\pm$ 0.022 \\
2016 May  03 & 0.038 $\pm$ 0.021 & 0.031 $\pm$ 0.025 \\
2016 Jul  13 & 0.047 $\pm$ 0.019 & 0.034 $\pm$ 0.025 \\
2017 May  05 & 0.025 $\pm$ 0.020 & 0.030 $\pm$ 0.024 \\
2017 Aug 11  & 0.028 $\pm$ 0.023 & 0.030 $\pm$ 0.026 \\
2017 Aug 24  & 0.025 $\pm$ 0.022 & 0.027 $\pm$ 0.025 \\
\enddata
\tablecomments{Both the distortion value and distortion uncertainty value reported here is the median value without edges.
Edges are defined when distortion or distortion uncertainty is larger than 0.3 pixel, which comes from the lack of sample stars on the edge.}
\end{deluxetable}

In this way, we derive the final local distortion maps for all non-\emph{06-14} AO epochs in arcsec. The distortion maps are converted into pixel coordinates for each image using the previously derived transformations.
The local distortion map for 2016 May 03 is shown in the first two panels of Figure \ref{fig:distmap} as an example.
We apply the local distortion to the stars' original position and add the distortion uncertainty in quadrature. 
The 3rd panel of Figure \ref{fig:distmap} shows how the residuals are reduced 
between the \textit{06-14} alignment propagated starlist and 2016 May 03 starlist after 
applying the distortion map. The median residual has been reduced 
from 0.06 pixel (black arrows) to 0.04 pixel (red arrows).

\begin{figure}[htp]
\plotone{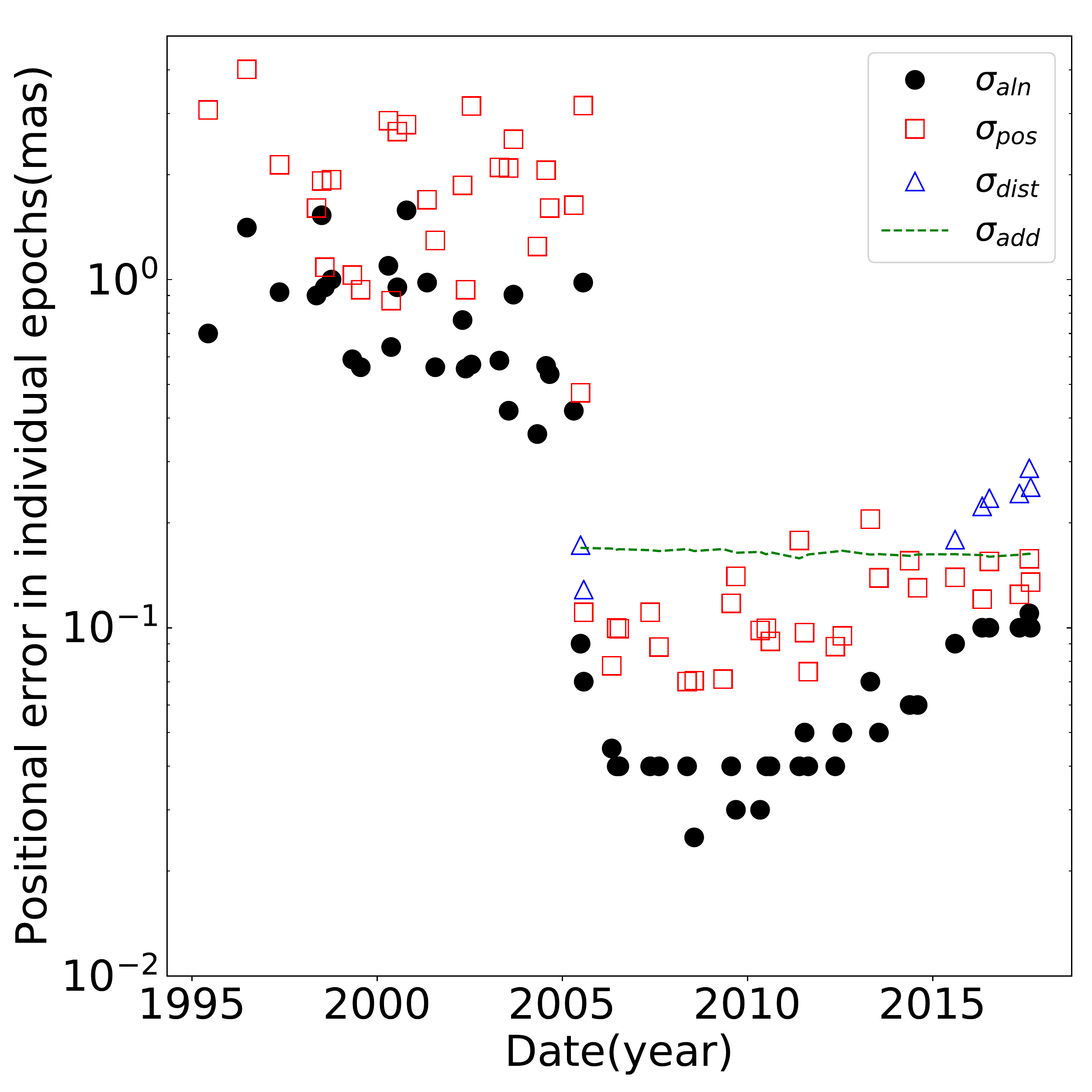}
\caption{The different types of positional errors in individual epochs.
Here the positional uncertainty is the median value for stars brighter than 16 magnitude and within 2$\arcsec$ from Sgr A*.
We only choose those stars because Speckle epochs have a smaller field of view and are much shallower (see details in \S\ref{sec:spe_align}).
Red open squares, black filled circles and Green dashed lines are \sigpos, \sigaln, \sigadd \ respectively (see details in \S\ref{sec:additive error}).
Blue open triangles show the distortion error added to non-06 epochs, $\sigma_{dist}$.
Notice that \sigpos \ and \sigaln \ exist for both Speckle and AO epochs, but \sigadd \ only exist for AO epochs (see \S\ref{sec:spe_align}) and $\sigma_{dist}$ only applies to epochs with local corrections.
}
\label{fig:err_epoch}
\end{figure}

Table \ref{tab:local_dist} summarizes the local distortion median value and typical uncertainties for all 8 non-06 epochs. 
To compare with the previously mentioned positional uncertainties in \S\ref{sec:additive error}, we plot the uncertainties as a function of time in Figure \ref{fig:err_epoch}.
From this plot, we can see AO epochs have reduced \sigpos \ and \sigaln \ by almost an order of magnitude relative to Speckle epochs.
For Speckle epochs, \sigpos \  is slightly larger than \sigaln \ and both contributes the total positional error.
For AO epochs, \sigadd\ and \sigpos\ are constant among epochs, but \sigaln \ increases with time after 2011, the reason being our reference epoch is 2009, so alignment transformation gets worse when further away from reference epoch.
For epochs with local distortion map, local distortion map error $\sigma_{dist}$ is slightly larger than other uncertainties.

\subsection{Adding Speckle Data}
\label{sec:spe_align}

\paragraph{New Holography Data}
In the new Holography analysis (see \S\ref{sec:starlist}), the stars' positional uncertainties are more accurately measured with bootstrapping, which captures unknown uncertainties from confusion or imperfect PSF.
So there is no need to add extra additive error.
The typical positional error for bright stars within 2$\arcsec$ from Sgr A* is plotted in Figure \ref{fig:err_epoch}.

\begin{figure}[htp]
\epsscale{1.1}
\plotone{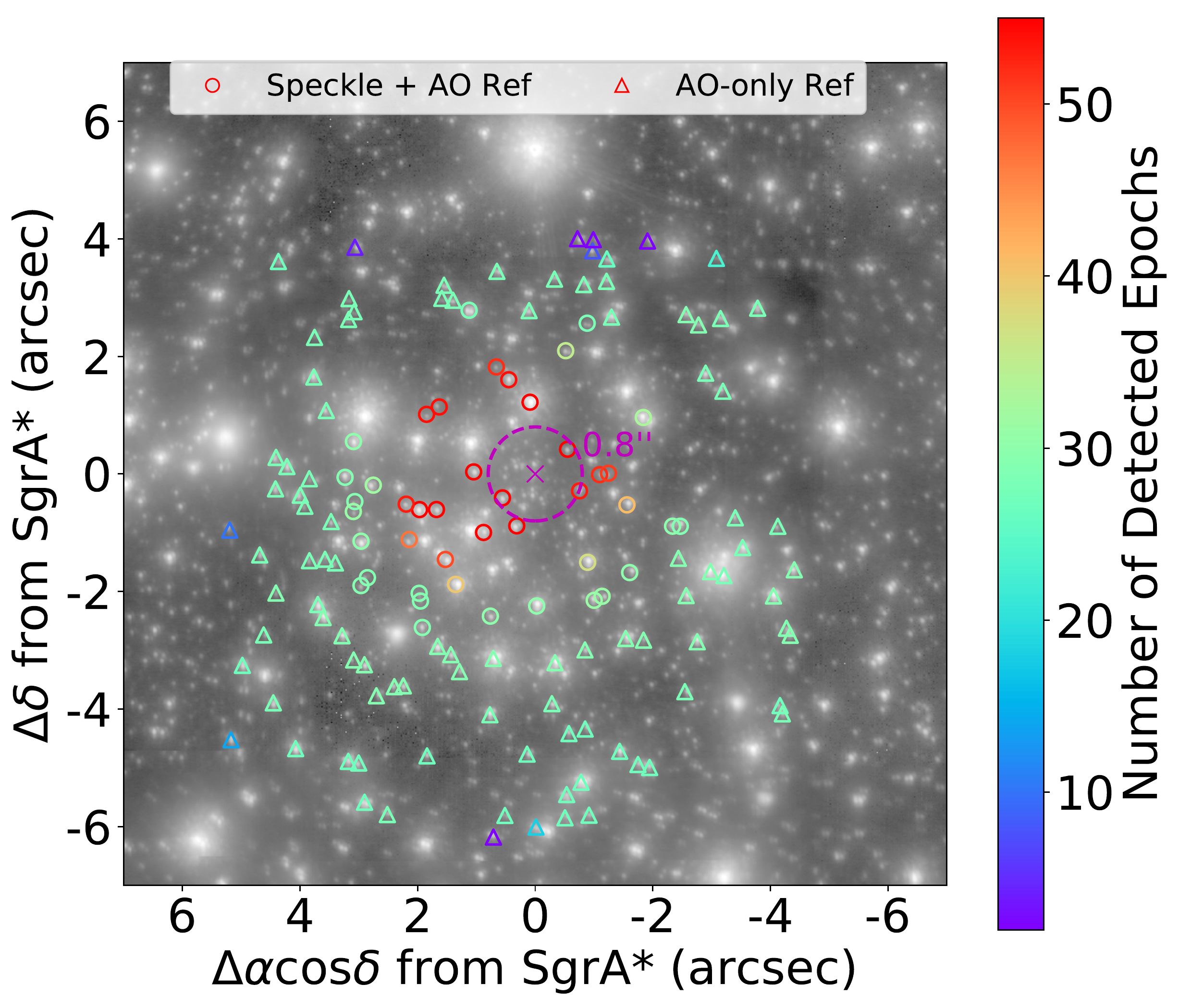}
\caption{The spatial distribution of \emph{reference stars} color coded with the number of epochs when they are used for calculating transformation. 
The triangles are AO only \emph{reference stars} while the circles are Speckle+AO \emph{reference stars}.
Most of the \emph{reference stars} out of 4$\arcsec$ are only used in less than 30 epochs, because those stars are not in Speckle's field of view, and are only detected AO epochs.
\emph{reference stars} are also required to be detected in more than 60\% of frames compared with IRS16C, in order to make sure they have reliable position measurements.
So there are a few stars in the outer region which are only used in less than 10 epochs.}
\label{fig:ref_dist}
\end{figure}

\paragraph{Reference Stars for Speckle}
From \S \ref{sec:ref_ao}, we have 141 \emph{reference stars} for AO epochs,
but the Speckle data has a smaller field of view and shallower detection limit. 
Therefore we make a radius cut of stars within 4$\arcsec $ from Sgr A*. 
This gives us 43 stars out of 141 stars. 
More importantly,  the Speckle images were not taken in stationary mode, so the field changed over night.
Stars on the edge of Speckle images will have less frame coverage compared with stars in the inner region, therefore they have relatively poor astrometric measurements.
To account for this effect, we require stars to be detected in more than 60\% of frames in Speckle epochs relative to IRS16C, which is one of the brightest and cleanest star in our field of view.
This 60\% criteria is lower compared to \cite{Boehle_2016} because new Holography has more frames.
The spatial distribution of AO and Speckle \emph{reference stars} is plotted in Figure \ref{fig:ref_dist}.

In conclusion, we have 141 \emph{reference stars} for AO epochs and 43 \emph{reference stars} for Speckle epochs. 
This is less than what \cite{Boehle_2016} used in her paper, mainly because we pursue the high quality of \emph{reference stars} over the quantity.
On average, we have 21 stars used as \emph{reference stars} for Speckle epochs and  133 stars for AO epochs, while \cite{Boehle_2016} has 61 stars for Speckle epochs and 230 stars for AO epochs.

\paragraph{Speckle Edge removal}
Since the number of frames that contribute to a given pixel near the edges of the Speckle images can be very low because of the field rotation,  stars would have poor astrometric measurements on the edge.
Therefore we decided to remove detections without enough frame coverage (less than 60\% frames relative to IRS16C) for Speckle epochs.
As a result, 3601 detections from 1020 stars were removed, which is almost 44.8\% of all stars over all Speckle epochs.

\subsection{Improved Matching in Crowded Region: Confusion Removal}
\label{sec:matching}
Up to this point, 2,700 stars are identified with a total of 56,061 measurements across 56 epochs.
However, some measurements are biased or incorrect due to mis-matches from stellar crowding and confusion.  
When matching starlists from different epochs, we require that a new measurement must fall within a radius of 40 mas of the predicted position.  
However, when two stars get too close to each other (e.g. within the first airy ring), \textit{StarFinder} cannot easily distinguish them.
In this case, only one source will be detected and the position for this source is biased as it is the flux-weighted average of the two stars. 
Fortunately, the probability of mis-matches for any one star changes with time as stars move past each other in projection; so the individual stars are not entirely lost if we can remove the biased epochs.

\begin{figure}[h]
	\plotone{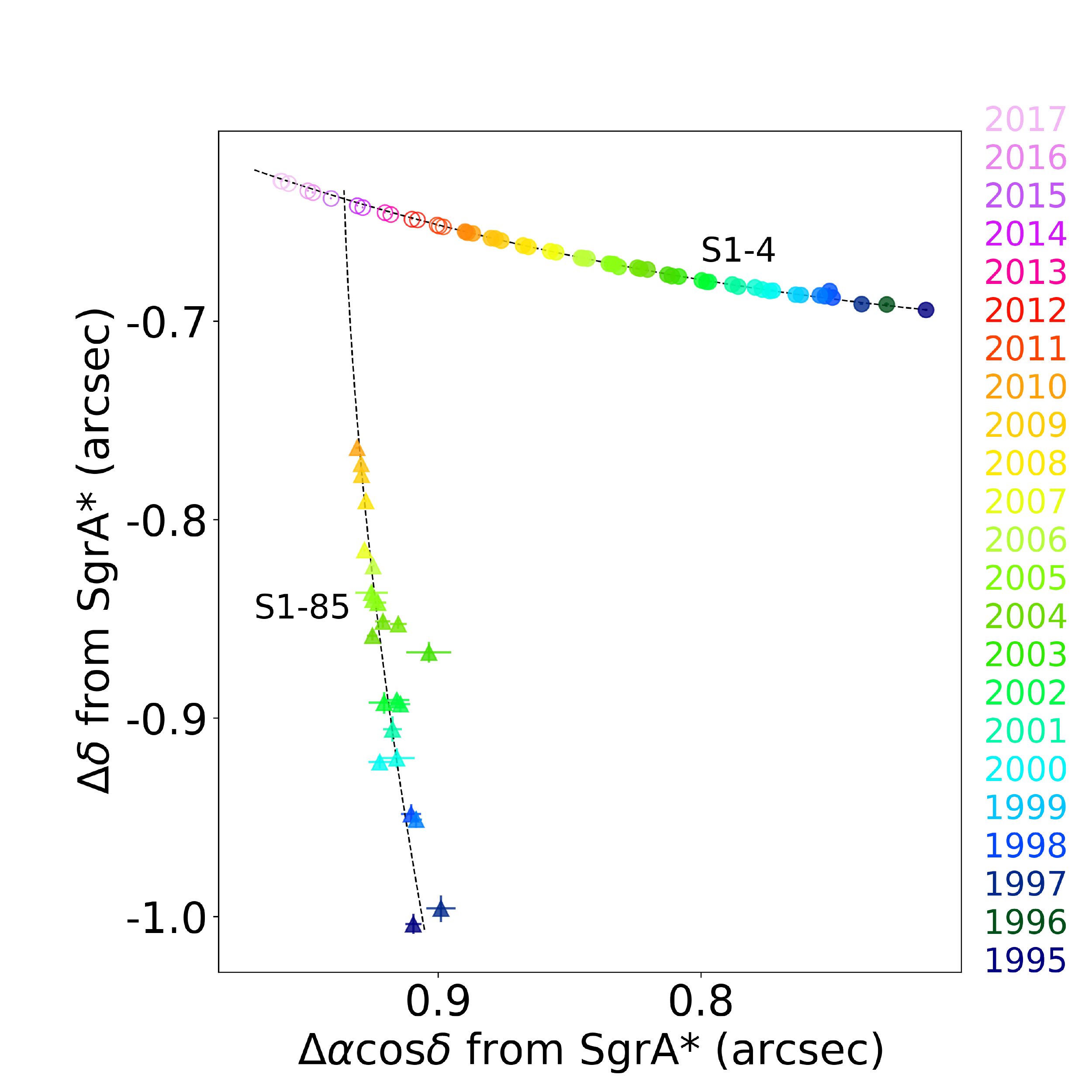}
    \caption{An example of how epochs with potential source confusion are removed: circles represent star S1-4 (K$'$=12.3)
                      and triangles represent star S1-85 (K$'$=15.3).
                     Different color shows stars' positions over time from 1995 to 2017.
                     The two stars move closer to each other.
                     From 1995 to 2010, S1-4 and S1-85 are separated enough to both be detected.
                     But after 2011,  they are too close to be distinguished and only S1-4 is detected.
                     In this case, we remove those detections for S1-4 after 2011 as shown in the open circles, because the position for S1-4 is biased by S1-85 in those epochs.}
	\label{fig:con}
\end{figure}

We choose to remove instances of confusion and mis-matching  using the following method:
For each star A, we find nearby stars that are within 100 mas. 
For every nearby star B, if star B is 5 mags fainter than star A or brighter, then star B is defined as a \emph{possible confusion source}. 
Notice that when star B is fainter than star A by more than 5 magnitude, the confusion from star B will not affect star A's position in a significant way.
Both star A and star B are required to be detected in more than 10 epochs for a reliable proper motion measurement.
For star A and its possible confusion sources, if only one source is detected  in some epoch where there should be two based on their proper motion prediction, this detection will be removed. 
Figure \ref{fig:con} shows an example on confusion removal.
In total, 5677 detections are removed because of confusion, affecting 751 stars, accounting for 10\% of the total detections.

\begin{figure*}[htp]
\plotone{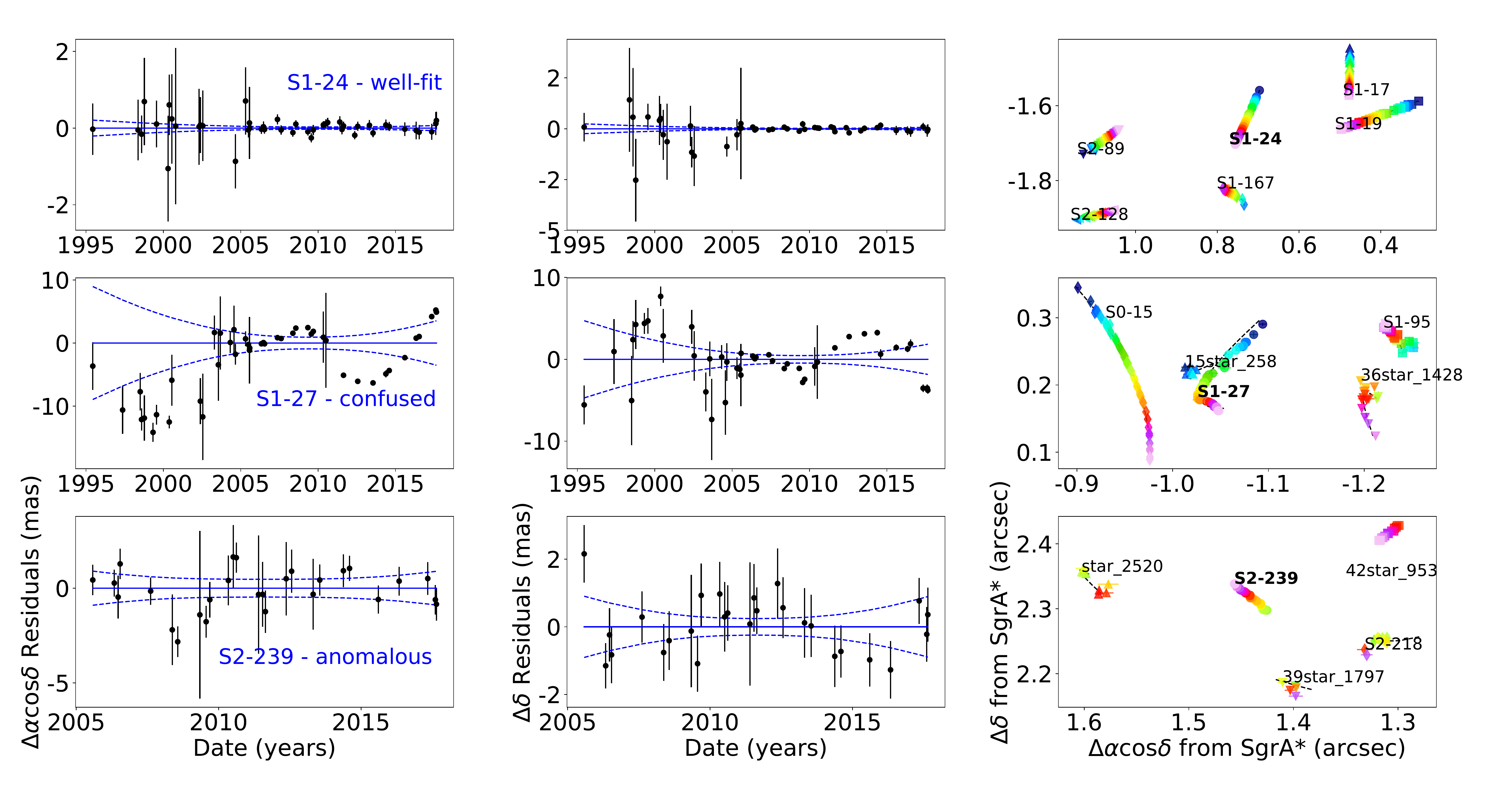}
\caption{
Examples of non-SMBH-acceleration sources from 3 different categories: ``well-fit'' (\emph{top} row), ``confused'' (\emph{middle} row) and ``anomalous'' (\emph{bottom} row), which is defined by how well the acceleration fit is and whether there are confusion stars nearby.
Here, the first two columns plot the residuals from acceleration fit in $\alpha\cos\delta$ and $\delta$ direction, and the third column plots the nearby stars.
      }
\label{fig:acc_cate}
\end{figure*}

\section{Proper Motions, Accelerations and Orbits}
\label{sec:poly}
With the well measured astrometric positions from \S\ref{sec:align}, we can measure the proper motion, acceleration and orbits for a final sample of 1148 stars.
In order to obtain a precise estimate of the proper motion, we require stars to be detected in more than 20 epochs (2/3 of all AO epochs), which gives us a sample of 1184 stars. 
In this sample, the median positional uncertainty is 2.37 mas for speckle epochs and 0.25 mas for AO epochs.
The positions over time are then used to fit a kinematic model for each star consisting of either a first-order (linear) polynomial, a second-order (acceleration) polynomial, or a full Keplerian orbit. 
We use a full orbit fit for the 33 stars within 0\farcs5 of the SMBH.
For the stars outside 0\farcs5, we first fit a second-order polynomial fit to derive the acceleration for all stars.
A jackknife is used to get a robust estimate of the acceleration uncertainty. 
We find 103 significant accelerating stars at the $>$ 5$\sigma$ level, which are then further divided into SMBH-acceleration sources and non-SMBH-acceleration sources in \S\ref{sec:acc}.
Then the remaining 1048 stars are fit with a first-order linear motion model.
The median velocity for those 1048 linear moving stars is 0.05 mas yr$^{-1}$ in each direction.

\subsection{SMBH-acceleration and Non-SMBH-acceleration sources}
\label{sec:acc}
The sample of 103 stars with significant ($>5 \sigma$) accelerations are analyzed to determine whether the best-fit accelerations are consistent with the expected acceleration from the SMBH (i.e. negative radial acceleration).
The accelerations are projected into the radial ($a_r$) and tangential ($a_t$) directions w.r.t. the SMBH.
Significant accelerating sources are defined when one of the following conditions is satisfied:
(1) radial acceleration is 5 sigma larger than its uncertainty: $|a_r/\sigma_{a_r}| > 5$.
(2) tangential acceleration is 5 sigma larger than its uncertainty: $|a_t/\sigma_{a_t}| > 5$.

Then, we define a sample of significant {\em SMBH-acceleration} sources when all of the following conditions are satisfied:
(1) significant negative acceleration:  $a_r/\sigma_{a_r}<-5$, 
(2) tangential acceleration is consistent with zero: $|a_t/\sigma_{a_t}|<3$, 
(3) negative radial acceleration is 3$\sigma$ smaller than allowed: $ (a_{r,max} - a_r)/\sigma_{a_r} <3 $, where $a_{r,max} = -GM/r^2$ is the maximum allowed radial velocity from the SMBH, M is the mass of the SMBH and r is the 2D projected distance. 
Since we don't have the line-of-sight distance, r will be the lower limit of the real 3D distance, so $a_{r,max}$ will be the upper limit of the allowed radial acceleration $a_r$.
These criteria result in 27 significant SMBH-acceleration sources. 

The remaining 76 accelerating sources are all {\em non-SMBH-acceleration} sources, including significant tangential accelerations, significant positive radial accelerations and too-large negative radial accelerations. 
The non-SMBH-acceleration sources are likely due to a number of factors including unrecognized confusion and binarity.
A potential binary candidate is analyzed in \S \ref{sec:binary}.

The large number of non-SMBH-acceleration sources suggests that there may be some contaminants in the SMBH-acceleration sample. 
To determine the degree of contamination, we check  all 103 stars by eye, and divide them into 3 categories.
1) ``well-fit": stars that show no time-coherent residuals from the acceleration model.
2) ``confused": stars that are not fit well by an acceleration model, but show potential confusion from neighboring stars.
3) ``anomalous": stars that are not fit very well by acceleration, and show no potential confusion around them. 

\begin{deluxetable}{ccc}
\caption{Accelerating Categories Summary} 
\label{tab:acc_cate}
\tablehead{\colhead{Category} & \colhead{SMBH-Accel} & \colhead{Non-SMBH-Accel}}
\startdata
 well-fit       & 24 & 54\\
 confused        & 3 & 13 \\
 anomalous      & 0 & 9 \\
\enddata
\end{deluxetable}

Figure \ref{fig:acc_cate} gives an example for each category from non-SMBH-acceleration sources. 
The first two columns are residuals from the acceleration fit and the third column shows the proper motion of nearby stars. 
To determine the category an accelerating source belongs to, we first look at the quality of the acceleration fit. 
In the figure, we can see S1-24 is fit well by an acceleration model given that the residuals are randomly distributed with time; so S1-24 is a ``well-fit'' non-SMBH-acceleration source. 
If a star is not fit well by an acceleration model, like S1-27 and S2-239, we consider alternative explanations: confusion from nearby stars, bad measurements, or other physical explanations such as astrometric wobble due to binarity or microlensing.  
To exclude confusion, we check whether there are nearby stars that are not detected in all epochs 
(note that in \S\ref{sec:matching}, confusion events are removed for cases when both stars are detected in more than 10 epochs, so stars detected in fewer epochs will be missed as potential confusion sources).
S1-27 has a nearby star 15star{\_}258 which is only detected in a few epochs, so S1-27 is potentially  confused with 15star{\_}258 when 15star{\_}258 is not detected\footnote{Notice that all stars in the final sample have names such as "S{\em RR}-{\em NN}", where {\em RR} is the radius this star belong to.
Stars which are not in our final sample have temporary names such as "{\em NN}star\_{\em NN}".}. 
In comparison, S2-239 is very isolated. 
Therefore, S1-27 is classified as a ``confused'' non-SMBH-acceleration source and S2-239 is an ``anomalous'' non-SMBH-acceleration source.

\begin{figure}
\plotone{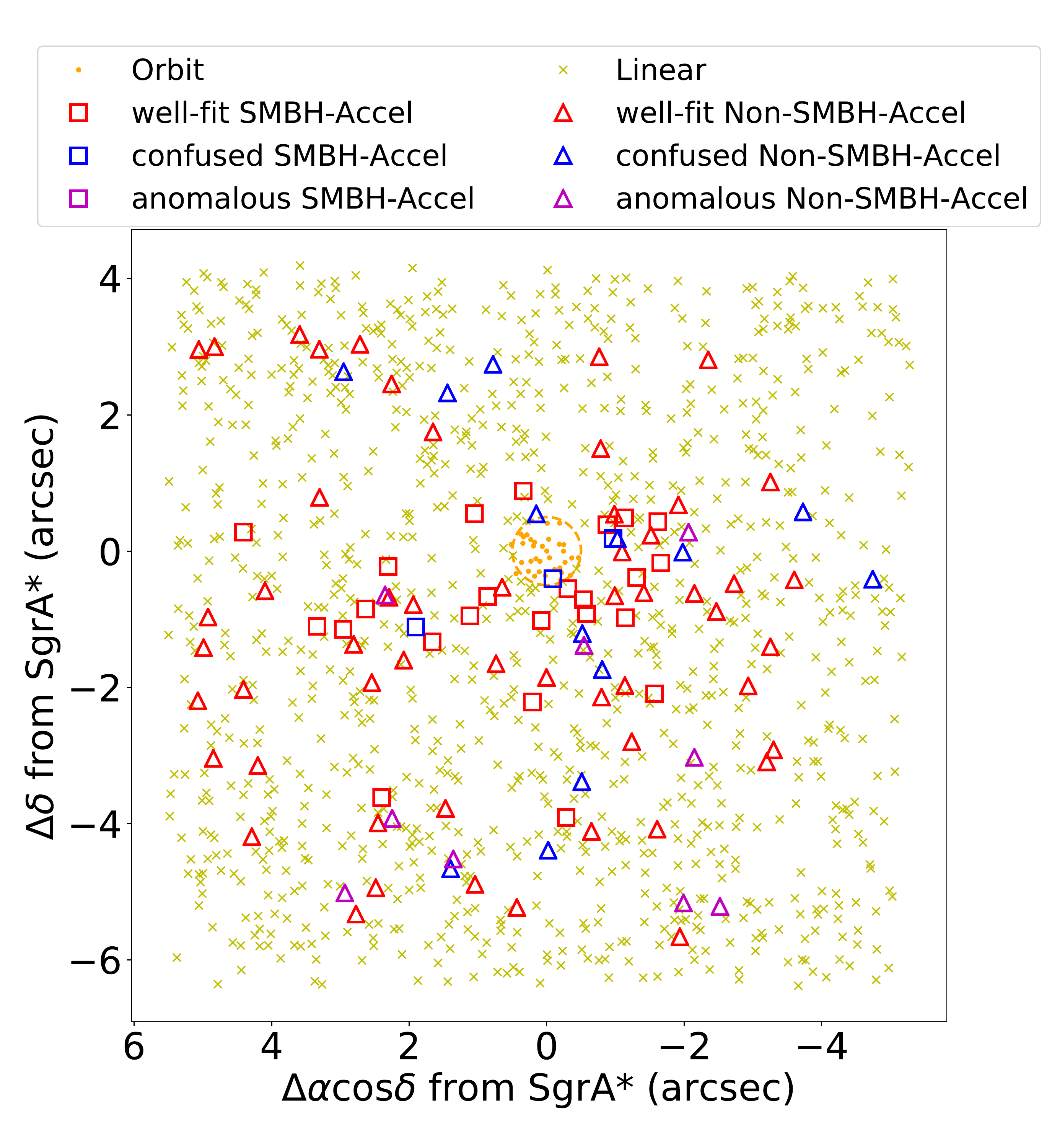}
\caption{The spatial distribution of our final sample: 33 orbital stars, 1048 linear moving stars, 27 significant SMBH-acceleration sources and 76 significant non-SMBH-acceleration sources, 
where the significant SMBH-acceleration and non-SMBH-acceleration sources are further divided into well-fit, confused, and anomalous as described in \S\ref{sec:acc}. 
Stars within 0\farcs5  from Sgr A* are orbital stars {\em (orange points)} and the orange circle shows the boundary of 0\farcs5.
Linearly moving stars are shown as {\em yellow crosses}.
Stars with significant acceleration (more than 5$\sigma$)  are then divided into SMBH-acceleration sources {\em (squares)} and non-SMBH-acceleration sources {\em (triangles)}.
These are further sub-divided into well-fit {\em (red)}, confused {\em (blue)}, and anomalous {\em (magenta)} stars. 
}
\label{fig:all_cate}
\end{figure}

In summary, Table \ref{tab:acc_cate} shows the number of accelerating sources in different categories.
Figure \ref{fig:all_cate} plots the spatial distribution of stars from: orbit(33), acceleration (103) or linear motion (1048), where acceleration stars are further  divided into the categories as showed in Table \ref{tab:acc_cate}.
The 24 well-fit SMBH-acceleration sources are particularly interesting and are explained in detail in \S\ref{sec:new_accel}.
The 76 non-SMBH-acceleration sources require further analysis, in which S0-27 is used as an example to explore the potential binarity in \S \ref{sec:binary}.

\section{Results} 
\label{sec:result}
The improvements in the astrometric methodology described in \S\ref{sec:align} deliver astrometry that is more accurate and more precise ($\sigma_V$ reduced by 40\%) compared to \cite{Boehle_2016}. 
A more detailed comparison is described in \S\ref{sec:align_assess}.
Based on this analysis, we detect 24 high-quality SMBH-acceleration stars, which are discussed in \S\ref{sec:new_accel}.
Finally, by fitting S0-2's orbit, we find the SMBH's position offset has been reduced by a factor of 2 in $\alpha\cos\delta$ direction and linear drift has been reduced by a factor of 4 in $\delta$  direction in \S\ref{sec:x0y0}.

\begin{figure}[htp]
\plotone{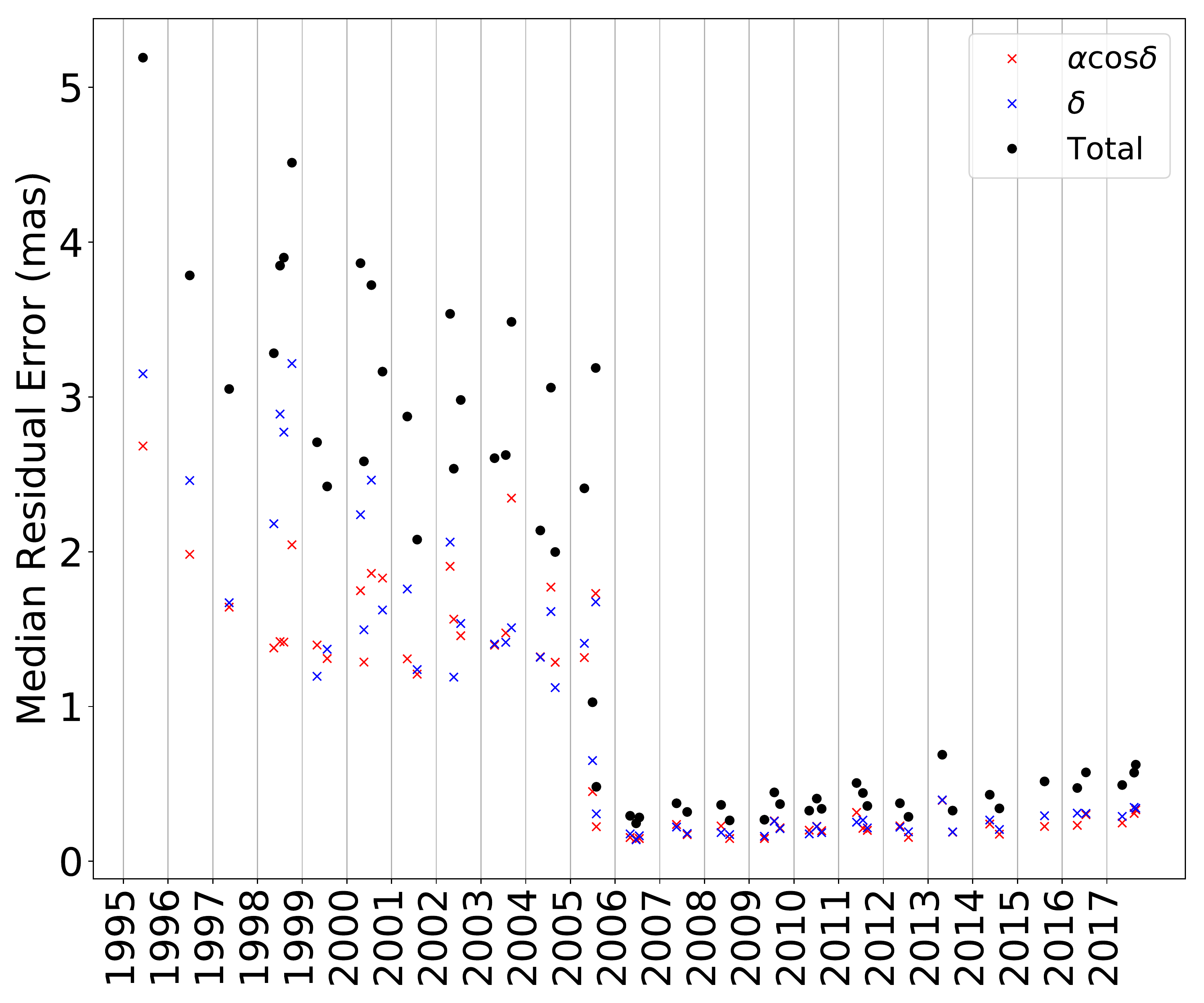}
\caption{The median residual from each stars' best-fit motion model for different epochs. 
Here, the model is either linear or acceleration based on which category the star belongs to.  
The sample plotted here includes stars outside of 0\farcs5, within 4$\arcsec$ from Sgr A*, and brighter than 16 magnitude in K$'$. 
The red and blue crosses show the median residual in the $\alpha\cos\delta$ and $\delta$ direction separately, while the black dot is the total residual. 
}
\label{fig:sigma_epoch}
\end{figure}

\begin{figure*}[htp]
\plotone{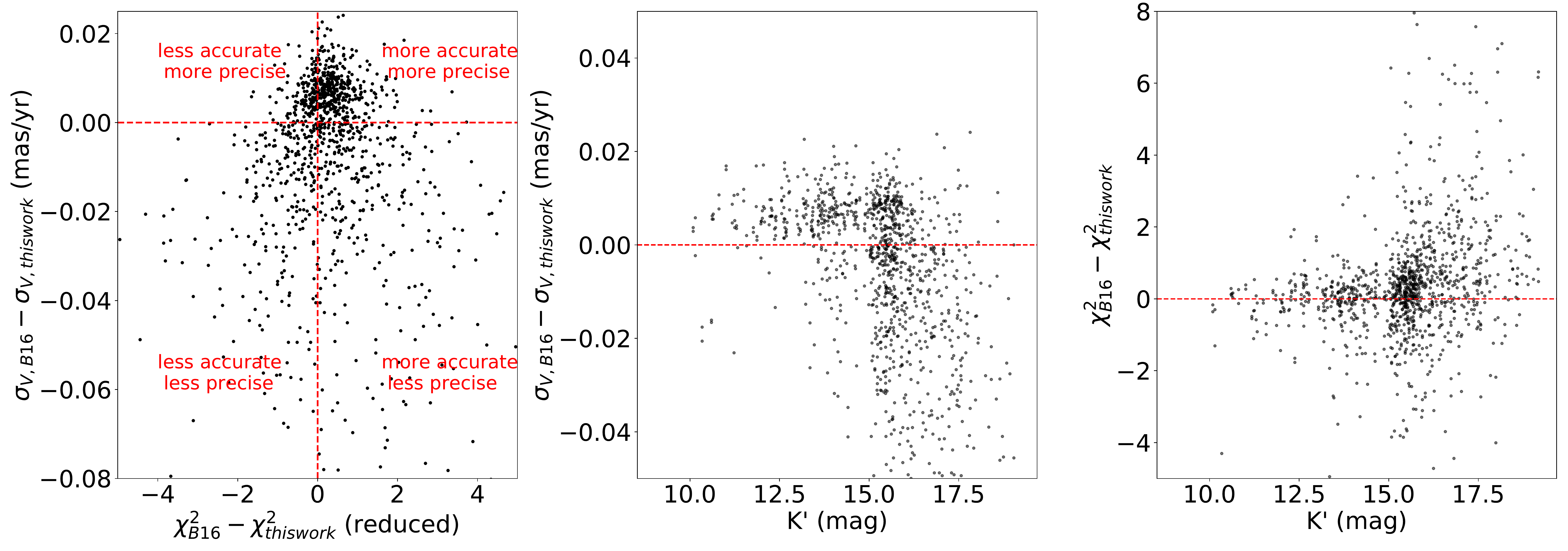}
\caption{The comparison of $\chi^2$ and $\sigma_V$ for linear moving stars between B16 and this work. 
A smaller $\chi^2$ indicates a more accurate measurement and a smaller $\sigma_V$ indicates a more precise measurement.
The \emph{left} panel plots the $\Delta_{\chi^2}$ and $\Delta_{\sigma_V}$ between B16 and this work.
The \emph{middle} and \emph{right} panel show how  $\Delta_{\chi^2}$ and $\Delta_{\sigma_V}$ depend on magnitude. 
}
\label{fig:chi2_ve}
\end{figure*}

\subsection{Improved Astrometry}
\label{sec:align_assess}

Stars are divided into 3 categories based on the order of the kinematic model needed to fit the on-sky positions: orbit, acceleration or linear motion (\S\ref{sec:poly}). 
To evaluate the aggregated goodness of the linear or acceleration fit,  Figure \ref{fig:sigma_epoch} shows the median fitting residual as a function of time. 
Given the complexity of matching in the dense region around the SMBH, the orbit stars within the central 0\farcs5 from Sgr A* are not included. 
Faint stars are more likely to be confused or biased in the measurement, so we only include stars brighter than 16 magnitude. These two cuts yield a sample of 553 stars that are fit by linear motion or acceleration models based on the categories they belong to. 
From the Figure \ref{fig:sigma_epoch}, we can see that the residuals from speckle epochs are 8 times larger than AO epochs, which is expected given the larger positional uncertainty in speckle epochs.
In fact, Speckle and AO epochs both have residuals around 1$\sigma$ relative to their positional uncertainty.

\begin{figure}[htp]
\plotone{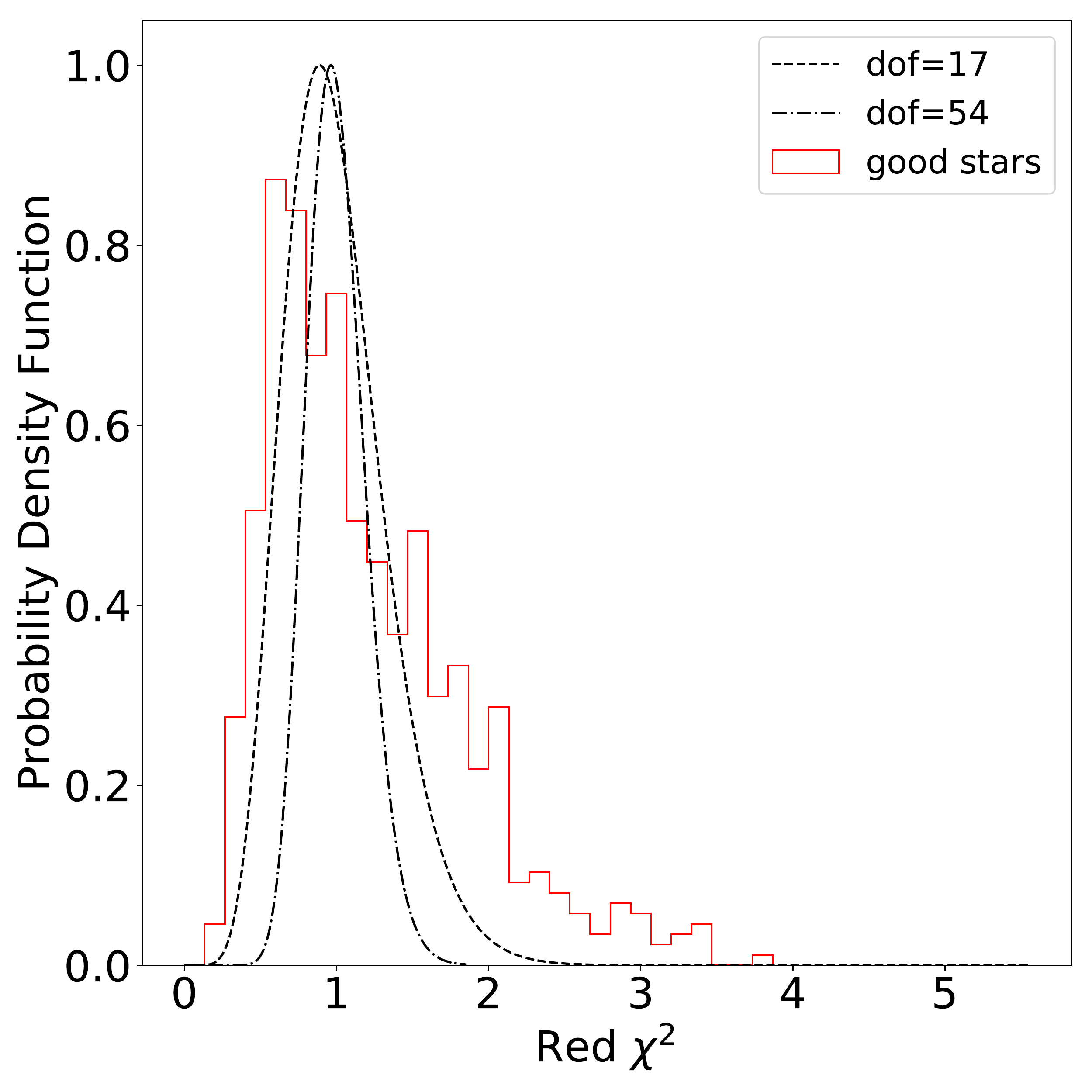}
\caption{The final reduced $\chi^2$ probability density function for \emph{good stars}.
Here reduced $\chi^2$ comes from linear/acceleration fit based on the category defined in \S \ref{sec:poly}.
Since stars are detected in different number of epochs and use different model fit (linear/acceleration), \emph{dof} would vary between stars.
To account for that effect, we plot the standard reduced $\chi^2$  distribution for both the minimum \emph{dof} and maximum \emph{dof} in black lines.
Reduced $\chi^2$ from \emph{good stars} (\S \ref{sec:additive error}) is plotted in red histogram.
}
\label{fig:chi2_tot}
\end{figure}

To evaluate the quality of our astrometry measurements, we also compare the $\chi^2$ distribution from linear fits and the distribution of velocity uncertainties, $\sigma_V$, with \cite{Boehle_2016} (referred as B16 here after).
To make a fair comparison, we require stars to be detected in more than 20 epochs in the B16 alignment, which gives us a total of 596 stars in the comparison sample. 
The left panel from Figure \ref{fig:chi2_ve} clearly shows the astrometry in this work is more precise (smaller $\sigma_V$) and more accurate (smaller $\chi^2$).
While the increased time baseline of our data set (up to 2013 vs. 2017) partially contributes to the increased measurement precision, the methodology changes in our work increase both the precision and accuracy. 
From the middle panel, it is clear that almost all stars brighter than 15 magnitude have a smaller $\sigma_V$ in our analysis.
The median $\sigma_{V}$ is reduced by 40\% from 0.017 mas yr$^{-1}$ to 0.010 mas yr$^{-1}$ in the $\alpha\cos\delta$ direction and from 0.019 mas yr$^{-1}$ to 0.010 mas yr$^{-1}$ in the $\delta$ direction for those bright stars.
For faint stars, larger $\sigma_V$ is expected due to their larger position uncertainties and will give a more robust uncertainty measurement.
From the right panel, it is apparent that, for stars of all magnitudes, our analysis typically yields smaller $\chi^2$ values.
In summary, the majority of the stars favors our work relative to B16.

Figure \ref{fig:chi2_tot} plots the final reduced $\chi^2$ distribution for the \emph{good stars} defined in \S\ref{sec:additive error}. 
328 stars are detected in the final sample among 352 \emph{good stars}.
With the improvements made on \S\ref{sec:align}, the final $\chi^2$ distribution for those 328 stars from 56 years of observation agrees with the predicted $\chi^2$ distribution.  

\subsection{New Accelerating Sources}
\label{sec:new_accel}

\begin{figure*}[htp]
\plotone{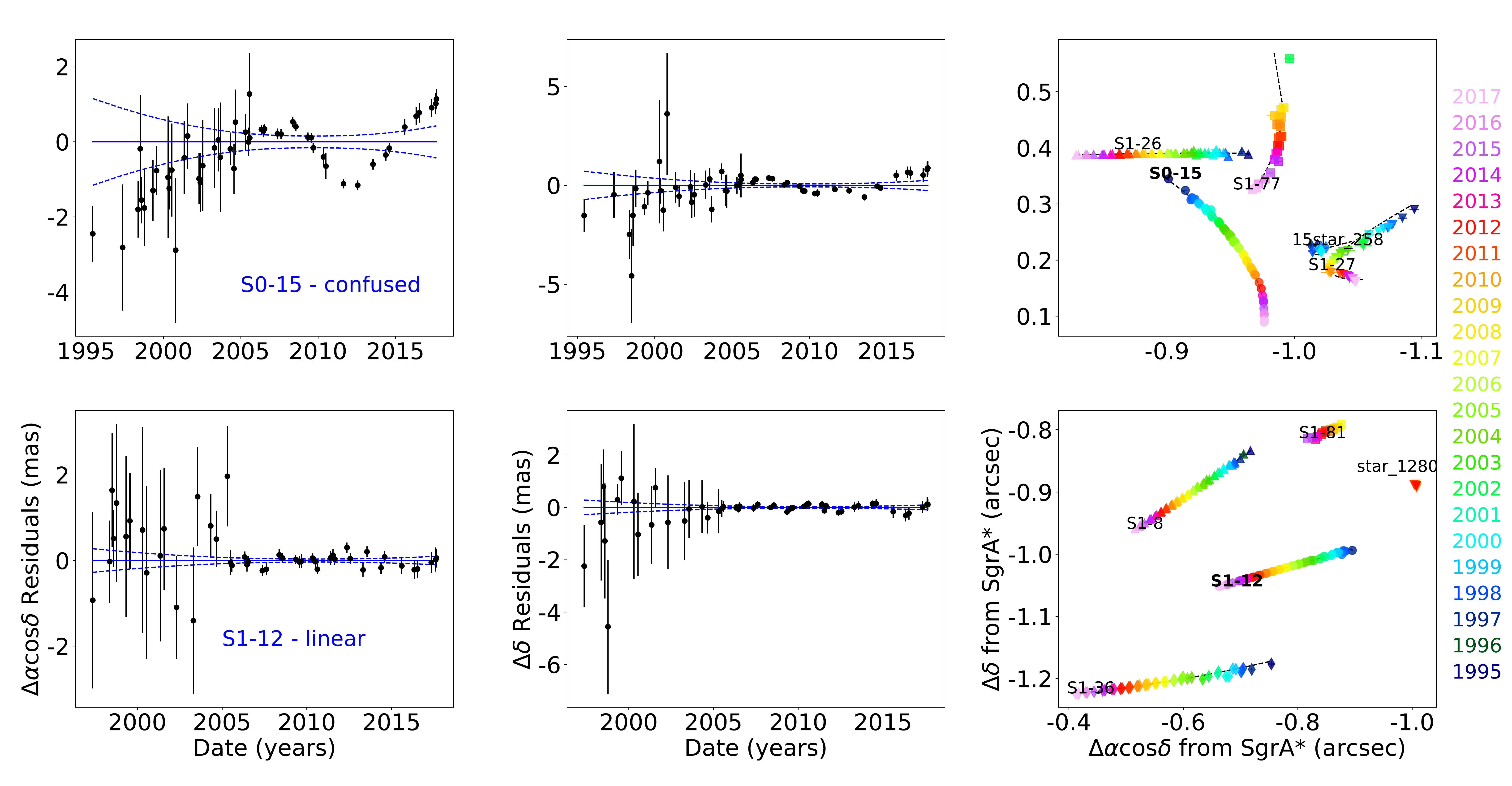}
\caption{The \emph{top}t row plots the acceleration fit for star S0-15 while the \emph{bottom} row plots the linear fit for star S1-12.
The first two columns are residual from acceleration/linear fit and the third panel shows the moving track of S0-15/S1-12 and nearby stars.
S0-15 and S1-12 are both accelerating sources in \cite{Yelda_2014}, but not in our analysis.
There is a clear correlation between the residuals and time in both $\alpha\cos\delta$ and $\delta$ direction for S0-15. 
Potential confusion might be the reason for this poor fit, as S0-15 and S1-27 gets very close after 2007. 
S1-12 is not accelerating by more than 5$\sigma$ in $a_r$ and its linear fitting is already satisfying, so we label S1-12 as a linear moving star. 
}
\label{fig:S015}
\end{figure*}

\begin{figure}[htp]
\plotone{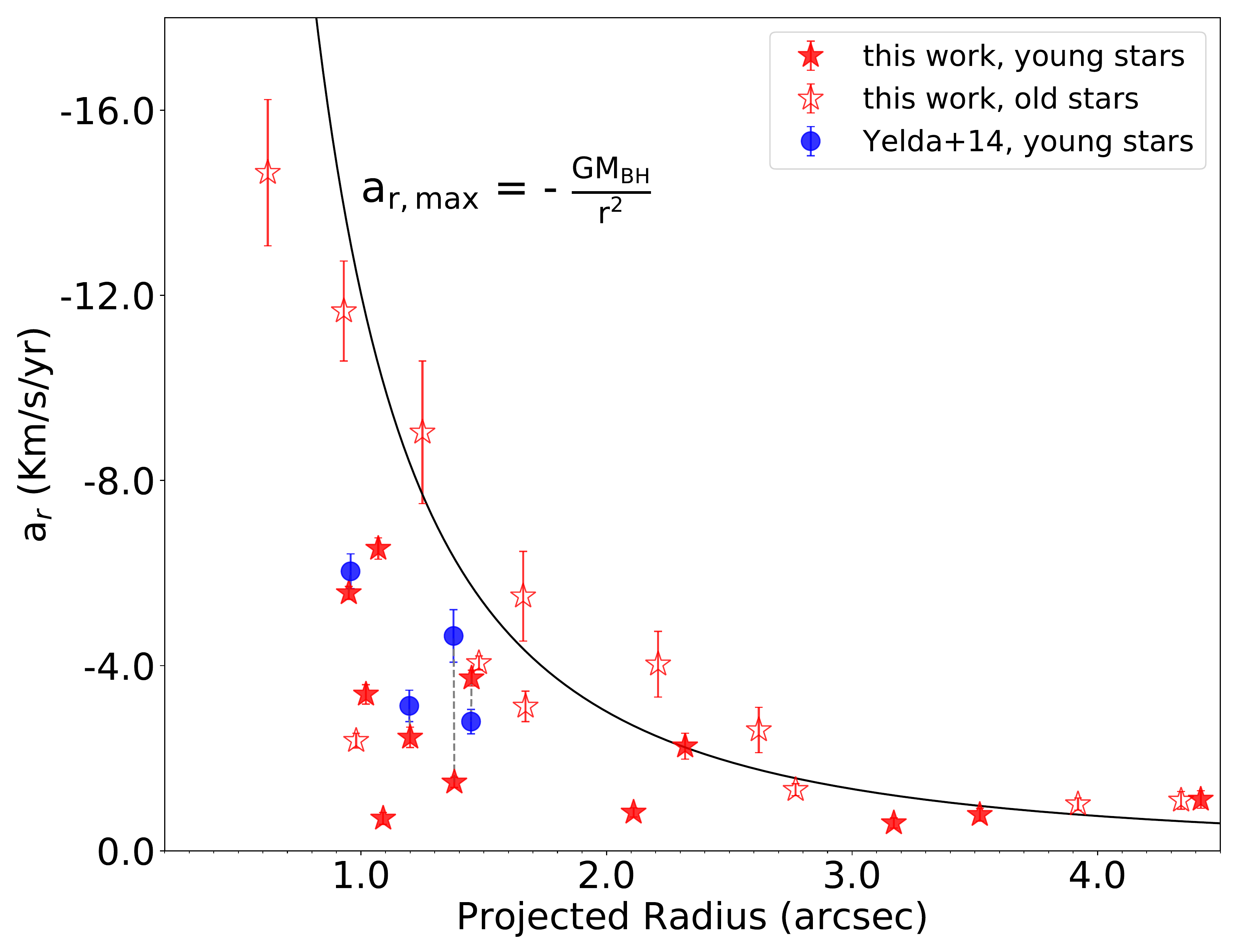}
\caption{The radial acceleration $a_r$ as a function of projected radius to Sgr A* for 24 well-fit SMBH-acceleration sources from this paper.
The filled red stars are young stars and the empty red stars are old stars. 
Four of them (S1-3,  IRS16C, S1-14 and  IRS16SW ) are previously published in \cite{Yelda_2014}, which are also plotted in blue circles. 
The 4 common stars are connected in dashed grey lines between our analysis and \cite{Yelda_2014}.
A solid black line shows the maximum allowed $a_r$ from Sgr A* calculated by $\mathrm{a_{r,max} = - GM_{BH}/{r^2}}$, where r is the projected distance to the SMBH.}
\label{fig:new_acc}
\end{figure}

From \S\ref{sec:acc}, we have identified 24 well-fit SMBH-acceleration sources outside of $r=$0\farcs5, which are summarized in Table \ref{tab:new_acc}.  
According to \cite{Do_2009}, 12 (50\%) of them are young stars. This is a large fraction considering that only 7\% of stars are young stars among our total 1184 stars.
This may suggest that young stars are very centrally concentrated compared to the old stars.  

\cite{Yelda_2014} published 6 young accelerating sources, 4 of which are also found accelerating in our analysis: S1-3,  IRS16C, S1-14 and  IRS16SW.
For those 4 young accelerating sources, the average $a_r$ uncertainty has been reduced by a factor of 2 from 0.39 $\mathrm{km s^{-1} yr^{-1}}$ to 0.16 $\mathrm{km s^{-1} yr^{-1}}$.
The radial acceleration for S1-3, IRS16C, and IRS16SW agree within 2$\sigma$ between our analysis and that of  \cite{Yelda_2014}.
S1-14 is discrepant by 3$\sigma$, but the $a_r$ uncertainties are very large in \cite{Yelda_2014}.
This large discrepancy comes from several aspects, among which the most important reasons are short time baseline and underestimated acceleration uncertainties in \cite{Yelda_2014}.
They only used data until 2011 (6 years AO observation), while we use data until 2017 (12 years AO observation), doubling the AO time baseline. 
Furthermore, the acceleration uncertainties are severely underestimated in \cite{Yelda_2014} because they did not use the jackknife method as we do in \S\ref{sec:poly}.
The other 2 stars that are accelerating in \cite{Yelda_2014} are S0-15 and S1-12. 
S0-15 is also accelerating in our sample, but because of its poor acceleration fit and the presence of nearby stars (Figure \ref{fig:S015}), we categorize it as a confused source. 
S1-12 is not accelerating by more than 5$\sigma$ in our analysis,  so it is characterized as a linear moving star.
\cite{Gillessen_2009_stars} also reported the following 5 accelerating sources: S0-70, S0-36, S1-3, S1-2 and S1-13, which are all also in our sample as listed in Table \ref{tab:new_acc}.

We plot our sample of 24 accelerating sources in Figure \ref{fig:new_acc}, among which 15 accelerating stars are reported for the first time.
Accelerations are detected for sources at 4'',  3 times further than previously published accelerations in \cite{Yelda_2014}.

\begin{deluxetable*}{lrrrrccc}
\tabletypesize{\footnotesize}
\tablecolumns{15} 
\tablewidth{0pt} 
\tablecaption{Significant SMBH-acceleration Sources}
\tablehead{
	\colhead{Name}      &
	\colhead{Mag}		&
    \colhead{Radius}		&
	\colhead{$a_r$}		&
    \colhead{$a_t$}        &
    \colhead{$a_r$ from \cite{Yelda_2014}}   &
    \colhead{$a_r$ from \cite{Gillessen_2009_stars}}   &
    \colhead{Young \tablenotemark{a}} \\
	\colhead{}      &
	\colhead{}		&
    \colhead{(arcsec)}		&
	\colhead{(km s$^{-1}$ yr$^{-1}$)}		&
    \colhead{(km s$^{-1}$ yr$^{-1}$)} &
    \colhead{(km s$^{-1}$ yr$^{-1}$)} &
    \colhead{(km s$^{-1}$ yr$^{-1}$)} &
    \colhead{}
            }
\startdata
     S0-70  & 17.8  &  0.62  & -14.65   $\pm$  1.58  &   2.33 $\pm$  1.70    & --                  &   -22.64 $\pm$ 3.02  &   --    \\
     S0-36  & 15.9  &  0.93  & -11.66   $\pm$  1.08  &   0.25 $\pm$  0.99    & --                  &   -13.21 $\pm$ 2.26  &   --    \\       
      S1-3  & 12.0  &  0.95  &  -5.57   $\pm$  0.14  &  -0.31 $\pm$  0.14    & -6.04 $\pm$  0.38   &   -2.64  $\pm$ 0.38  &   y     \\       
     S1-26  & 15.4  &  0.98  &  -2.38   $\pm$  0.16  &   0.12 $\pm$  0.16    & --                  &   --                 &   --    \\       
      S1-2  & 14.6  &  1.02  &  -3.38   $\pm$  0.21  &   1.03 $\pm$  0.36    & --                  &   -1.89  $\pm$ 0.38  &   y     \\       
      S1-4  & 12.3  &  1.07  &  -6.53   $\pm$  0.23  &  -0.37 $\pm$  0.21    & --                  &   --                 &   y     \\       
      S1-8  & 14.0  &  1.09  &  -0.70   $\pm$  0.13  &  -0.11 $\pm$  0.12    & --                  &   --                 &   y     \\       
    IRS16C  &  9.9  &  1.20  &  -2.45   $\pm$  0.22  &   0.35 $\pm$  0.26    & -3.13 $\pm$  0.34   &   --                 &   y     \\       
     S1-92  & 16.6  &  1.25  &  -9.04   $\pm$  1.54  &   1.01 $\pm$  1.36    & --                  &   --                 &   --    \\       
     S1-14  & 12.6  &  1.38  &  -1.48   $\pm$  0.11  &   0.27 $\pm$  0.13    & -4.64 $\pm$  0.57   &   --                 &   y     \\       
   IRS16SW  & 10.0  &  1.45  &  -3.73   $\pm$  0.17  &   0.08 $\pm$  0.17    & -2.79 $\pm$  0.26   &   --                 &   y     \\       
     S1-13  & 13.9  &  1.48  &  -4.06   $\pm$  0.15  &   0.40 $\pm$  0.15    & --                  &    -2.64 $\pm$  0.38 &   --    \\
     S1-47  & 15.5  &  1.66  &  -5.50   $\pm$  0.97  &   0.32 $\pm$  0.58    & --                  &   --                 &   --    \\ 
     S1-51  & 14.9  &  1.67  &  -3.12   $\pm$  0.33  &  -0.42 $\pm$  0.33    & --                  &   --                 &   --    \\ 
      S2-6  & 11.8  &  2.11  &  -0.83   $\pm$  0.10  &   0.11 $\pm$  0.10    & --                  &   --                 &   y     \\       
    S2-127  & 15.7  &  2.21  &  -4.03   $\pm$  0.71  &   2.04 $\pm$  1.03    & --                  &   --                 &   --    \\       
     S2-22  & 12.8  &  2.32  &  -2.26   $\pm$  0.28  &  -0.02 $\pm$  0.56    & --                  &   --                 &   y     \\       
    S2-219  & 15.8  &  2.62  &  -2.61   $\pm$  0.49  &   0.40 $\pm$  0.51    & --                  &   --                 &   --    \\       
     S2-75  & 14.3  &  2.77  &  -1.32   $\pm$  0.13  &  -0.33 $\pm$  0.15    & --                  &   --                 &   --    \\       
      S3-5  & 16.1  &  3.17  &  -0.60   $\pm$  0.11  &  -0.15 $\pm$  0.16    & --                  &   --                 &   y     \\
     S3-10  & 13.6  &  3.52  &  -0.78   $\pm$  0.14  &   0.34 $\pm$  0.14    & --                  &   --                 &   y     \\
    S3-370  & 13.5  &  3.92  &  -1.01   $\pm$  0.13  &   0.14 $\pm$  0.11    & --                  &   --                 &   --    \\       
    S4-139  & 14.4  &  4.34  &  -1.09   $\pm$  0.19  &  -0.20 $\pm$  0.20    & --                  &   --                 &   --    \\ 
    S4-169  & 13.5  &  4.42  &  -1.11   $\pm$  0.19  &   0.45 $\pm$  0.24    & --                  &   --                 &   y     
\enddata
\tablenotetext{a}{Young stars are published in \cite{Do_2009}.}
\label{tab:new_acc}
\end{deluxetable*}

\begin{figure}[htp]
\plotone{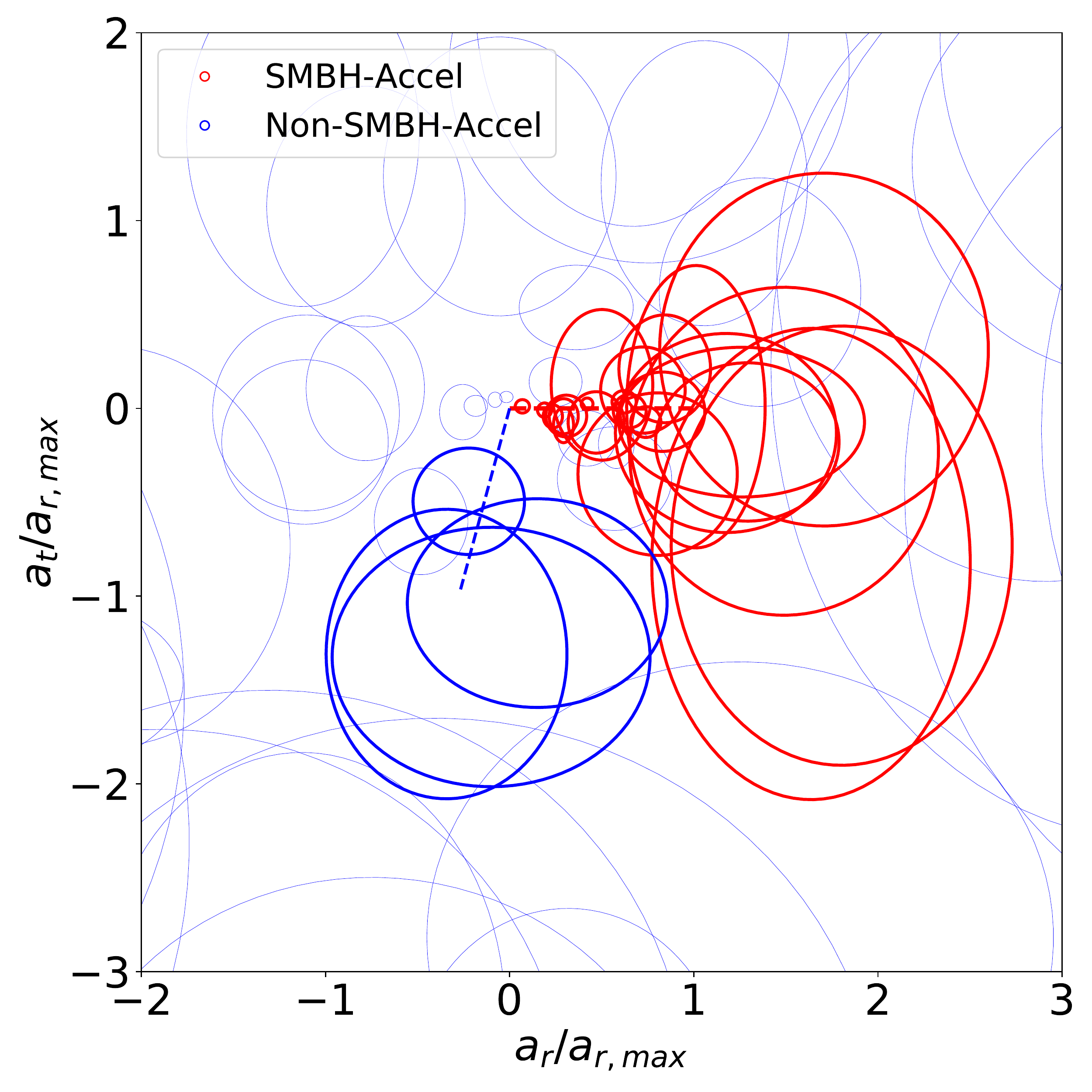}
\caption{Radial ($a_r$) and tangential ($a_t$) accelerations are used to determine how the non-SMBH-acceleration sources contaminate the SMBH-acceleration sample.
Each star's ellipse is centered on  [$a_r$/$a_{r,max}$,  $a_t$/$a_{r,max}$] with the semi-major axis of  3$\sigma_{a_r}$/$a_{r,max}$ and  3$\sigma_{a_t}$/$a_{r,max}$}. 
SMBH accelerations ({\em red ellipses}) will intersect with the red dashed line segment of length 1.
To determine the non-SMBH-acceleration contaminants, we draw segments with random orientations, originating from [0,0] with length of 1 and calculate how many non-SMBH-acceleration sources intersect with the segment, where non-SMBH-acceleration sources are plotted in blue ellipses.
A random example is shown as a blue dashed line that intersects 5 non-SMBH-acceleration sources (thick blue ellipses).

\label{fig:arat}
\end{figure}

Since non-SMBH-acceleration sources could  exist in all directions, we determine how many potentially non-SMBH-acceleration sources exist in our 24 SMBH-acceleration sample. 
First, we define well-fit SMBH-acceleration sources as those where the star's tangential acceleration agrees with 0 within 3$\sigma$: $-3\sigma_{a_t} < a_t < 3\sigma_{a_t}$ and the star's radial acceleration is between 0 and $a_{r,max}$ within 3 sigma: $(a_{r,max} - 3\sigma_{a_r}) < a_r < 3\sigma_{a_r}$.
In the 2D space of $a_t$ vs. $a_r$, a star in the SMBH-acceleration sample would have a $1\sigma$ error ellipse that encloses $a_t=0$ and would intersect with a line going from $a_{r,max}$ to $a_r=0$. 
Since $a_{r,max}$ varies between different stars, we divided $a_r$ and $a_t$ by $a_{r,max}$ to normalize them.

Figure \ref{fig:arat} shows $a_t$/$a_{r,max}$ versus $a_r$/$a_{r,max}$, where each stars is drawn as an ellipse with 
the semi-major axis of 3$\sigma_{a_t}$/$a_{r,max}$ and 3$\sigma_{a_r}$/$a_{r,max}$ respectively.
The SMBH-acceleration stars intersect the segment on the horizontal axis from 0 to 1 and are shown in red, and the non-SMBH-acceleration stars are shown in blue.
To determine the potential contaminants from non-SMBH-acceleration stars, we randomly draw segments with different orientations originating from (0,0) and with length of 1. 
An example is showed in the bold blue ellipses. 
The average number of intersected ellipse from non-SMBH-acceleration sources gives the contamination rate. 
We find that among our 24 SMBH-acceleration sources, 3.5$\pm$1.1 could potentially come from non-SMBH-acceleration contaminants.

To determine if the contaminants are correlated with radius, we perform similar tests for stars within 2\farcs5 and beyond 2\farcs5 Sgr A*.
Of the 17 SMBH-acceleration sources within 2\farcs5, we find that 2.8$\pm$1.2 may come from non-SMBH-acceleration contaminants, or 16\%.
Of the 7 SMBH-acceleration sources beyond 2\farcs5, we find that  0.6$\pm$0.8 may come from non-SMBH-acceleration contaminants, or 9\%.
Therefore the contamination rate is not strongly radially dependent.
Unfortunately, we cannot identify the specific contaminated sources at this time and future observations are needed.

\subsection{Systematic Bias for Sgr A*}
\label{sec:x0y0}
\begin{figure}
    \includegraphics[width=0.4\textwidth]{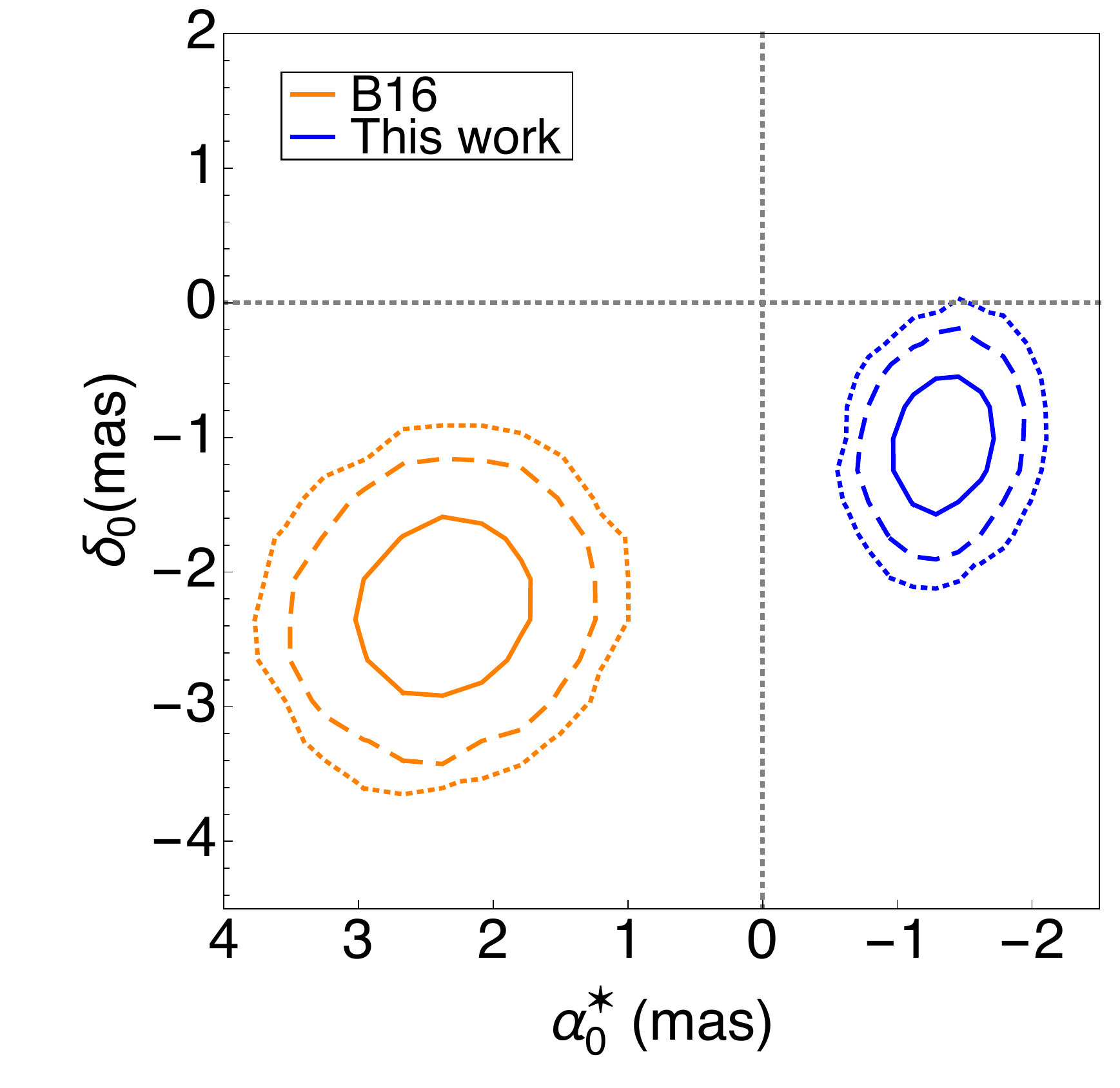}
    \includegraphics[width=0.4\textwidth]{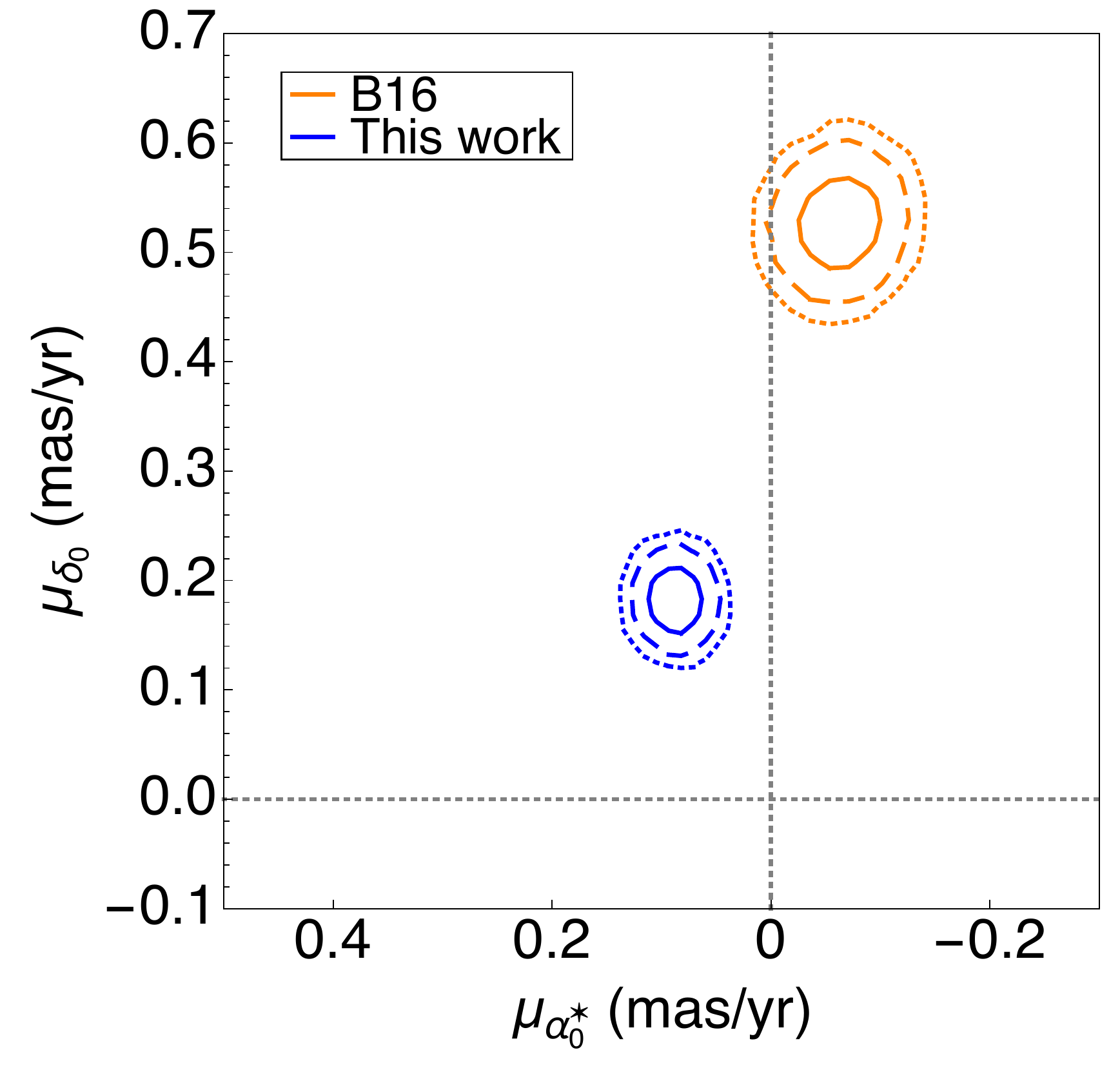}
    \caption{The posterior joint probability distributions of the position ($\alpha^{\ast}_0$, $\delta_0$) and proper motion ($\mu_{\alpha^{\ast}_0}$, $\mu_{\delta_0}$) of the SMBH based on S0-2's orbit fit.
    The upper panel shows the SMBH position and the lower panel shows the SMBH's linear drift.
    The orange contour is from \cite{Boehle_2016}, while the blue contour is from our work.
    The contours show 1$\sigma$, 2$\sigma$, and 3$\sigma$ uncertainties.
    Our cross-epoch alignment reduces the SMBH offset and linear drift in both directions.}
    \label{fig: x0y0}
\end{figure}

One of the goals of the methodology developed in this work is to construct a more stable reference frame for imaging observations of the Galactic Center. 
Currently, systematic uncertainties arising from the construction of the reference frame are assessed using orbital fits of short-period stars 
by including the astrometric position and velocity of the central SMBH as free parameters in the fit. 
A fit using S0-2's observations in \citet{Boehle_2016} has shown an offset in the position of the SMBH of 2.5 mas and a linear drift of 0.55 mas/yr. 
In this section, we assess the improvements induced by our new methodology on S0-2's orbital fit. 
We use astrometric observations from the cross-epoch alignment presented in \S\ref{sec:align}. 
In addition, we use S0-2's radial velocity measurements obtained using spectroscopic observations reported in \cite{Ghez_2003}, \cite{Ghez_2008}, \cite{Boehle_2016}, \cite{Chu_2018} and similar measurements from the VLT reported in \cite{Eisenhauer_2003}, \cite{Eisenhauer_2005}, \cite{Gillessen_2009}, \cite{Gillessen_2017} (a summary of all S0-2's radial velocity measurements can be found in \cite{Chu_2018}). 
The orbital fits are performed using Bayesian inference with the \texttt{MultiNest} sampler \citep{Feroz_2008, Feroz_2009}. 
The model used for the fit includes 13 parameters: 
the mass of the central SMBH, the distance to our Galactic Center $R_0$, the positions ($\alpha^{\ast}_0$ and $\delta_0$) 
and velocities ($\mu_{\alpha^{\ast}_0}$, $\mu_{\delta_0}$ and $v_{z_0}$) of the SMBH and the six orbital parameters of S0-2. 
Hereafter we will use $\alpha^{\ast}$ to represent $\alpha\cos\delta$.

Figure \ref{fig: x0y0} plots the posterior joint probability distributions of $\alpha^{\ast}_0$ and $\delta_0$ and of $\mu_{\alpha^{\ast}_0}$ and $\mu_{\delta_0}$  obtained using these observations. 
For comparison,  we also present the posterior probability distributions obtained using S0-2 astrometry from \citet{Boehle_2016}. 
Our new alignment methodology reduces the SMBH offset in both direction by a factor 2. 
The linear drift is also severely reduced in the $\delta$-direction by a factor 4. 
This analysis shows that the new methodology to align the different astrometric observations presented in this work and in \cite{Sakai_2019} improves the quality of the orbital fit of S0-2 significantly.

\section{Discussion}
\label{sec:discuss}

\subsection{Clockwise Disk of Young Stars}
\label{sec:young stars}
The age of the young stars (4-6 Myr) around the SMBH is much smaller than the relaxation time scale ($\sim$ 1 Gyr) in the GC, so their dynamical structure will greatly help us distinguish different star formation mechanisms \citep{Alexander_2017}.
Observations have found that around 20\% of the young stars move in a well-defined clockwise disk \citep{Levin_2003, Paumard_2006, Bartko_2009, Lu_2009, Yelda_2014}, while the rest off-disk stars are more randomly distributed, which may suggest an in-situ formation theory.
This 20\% fraction might only be a lower limit because of stellar binaries \citep{Naoz_2018}.
An accurate disk membership derivation is required to divide disk and off-disk stars correctly, and later compare useful properties, like the initial mass function, between them.

The disk membership is derived from the standard Keplerian orbital elements, including 6 kinematic variables ($\alpha^{\ast}, \delta, z, \mu_{\alpha^{\ast}}, \mu_{\delta}, v_z$).
\cite{Yelda_2014} assigned the disk membership for 116 young stars using data from 1995 to 2011. 
With a much longer time baseline from 1995 to 2017, which doubles the AO observation time compared to \cite{Yelda_2014}, and an improved astrometric analysis, our work will give a much more accurate estimate of the Keplerian orbital parameters and thus improve the disk membership assignment.

Among the 6 kinematic variables, only the line of sight velocity $v_z$ comes from the spectroscopic measurements, while the rest 5 parameters all come from the astrometric measurement.
The projected position ($\alpha^{\ast}$, $\delta$) and proper motion ($\mu_{\alpha^{\ast}}$, $\mu_{\delta}$) can be directly derived from \S\ref{sec:poly}.
For 48 linear moving stars which are reported both in \cite{Yelda_2014} and our final sample, the median velocity uncertainty is reduced  from 0.073 mas yr$^{-1}$ to 0.019 mas yr$^{-1}$, by almost a factor of 4.
The most difficult part is to measure the line-of-sight distance ($z$).
Fortunately, the absolute value of line-of-sight distance can be derived using Eq.~9 from \cite{Yelda_2014},  if we can measure significant $a_r$. 
Even if we do not have significant measurements of $a_r$, stars with 3$\sigma$ acceleration upper limits smaller than $a_{r,max}$ can provide lower limits on the line-of-sight distance. 
From \S\ref{sec:new_accel}, we have already detected 12 significant $a_r$ measurements for young stars, 2 times more than \cite{Yelda_2014}, and our acceleration uncertainty is reduced by a factor of 2. 
Therefore the 5 kinematic parameters ($\alpha^{\ast}$, $\delta$, $\mu_{\alpha^{\ast}}$, $\mu_{\delta}$, $z$) can all be better estimated with our improved astrometry.

\subsection{Potential Binaries}
\label{sec:binary}

\begin{deluxetable}{@{\extracolsep{4pt}}lcc}
\tabletypesize{\footnotesize}
\caption{Linear Model}
\label{tab:linfit}
\tablecolumns{3} 
\tablewidth{0pt}
\tablehead{
\colhead{Params} & \colhead{Prior}	& \colhead{Best Fit} 
}
\startdata
$\mu_{\alpha^{\ast}}$  (mas/yr)  &  flat prior in [0,1]         &  0.56  $\pm$ 0.01 \\ 
$\mu_{\delta}$  (mas/yr)         &  flat prior in [2,4]         &  3.28  $\pm$ 0.01 \\ 
$\alpha^{\ast}_0$  (\arcsec)     &  flat prior in [0.12, 0.15]  &  0.1396 $\pm$ 0.0002 \\ 
$\delta_0$  (\arcsec)            &  flat prior in [0.45, 0.50]  &  0.4808 $\pm$ 0.0003  
\enddata
\end{deluxetable}

\begin{deluxetable}{@{\extracolsep{4pt}}lcc}
\tabletypesize{\footnotesize}
\tablecolumns{3} 
\tablewidth{0pt}
\caption{Linear + Binary Model}
\label{tab:bifit}
\tablehead{
\colhead{Params} & \colhead{Prior}	& \colhead{Best Fit} 
}
\startdata
$\mu_{\alpha^{\ast}}$  (mas/yr)         &  flat prior in [0,1]                    &  0.61   $\pm$ 0.01 \\ 
$\mu_{\delta}$  (mas/yr)                &  flat prior in [2,4]                    &  3.15   $\pm$ 0.02 \\ 
$\alpha^{\ast}_0$  (\arcsec)            &  flat prior in [0.12, 0.15]             &  0.1388 $\pm$ 0.0003 \\ 
$\delta_0$  (\arcsec)                   &  flat prior in [0.45, 0.50]             &  0.4834 $\pm$ 0.0004\\ 
$\omega$  (degree)                      &  flat prior in [0, 360]                 &  42     $\pm$ 14  \\ 
                                        &                                         &  222    $\pm$ 16 \tablenotemark{*}\\
$\Omega$  (degree)                      &  flat prior in [0, 360]                 &  136    $\pm$ 3  \\ 
                                        &                                         &  316    $\pm$ 3 \tablenotemark{*}\\
$i$  (degree)                           &  flat prior in P(i) = sin(i) in [0,180] &  111    $\pm$ 3 \\ 
$e$                                     &  flat prior in P(e) = e in [0,1]        &  0.40   $\pm$ 0.08 \\ 
$t_p$  (year)                           &  flat prior in [2000, 2020]             &  2010.2 $\pm$ 0.4  \\ 
$P$  (year)                             &  flat prior in [0, 30]                  &  12.7   $\pm$ 0.6 \\ 
$a$  (mas)                              &  flat prior in [0, 4]                   &  1.55   $\pm$ 0.08 \\ 
\enddata
\tablenotetext{*}{$\omega$ and $\Omega$ have two solutions because they are degenerate with each other.}
\end{deluxetable}

\begin{figure*}[htp]
\plotone{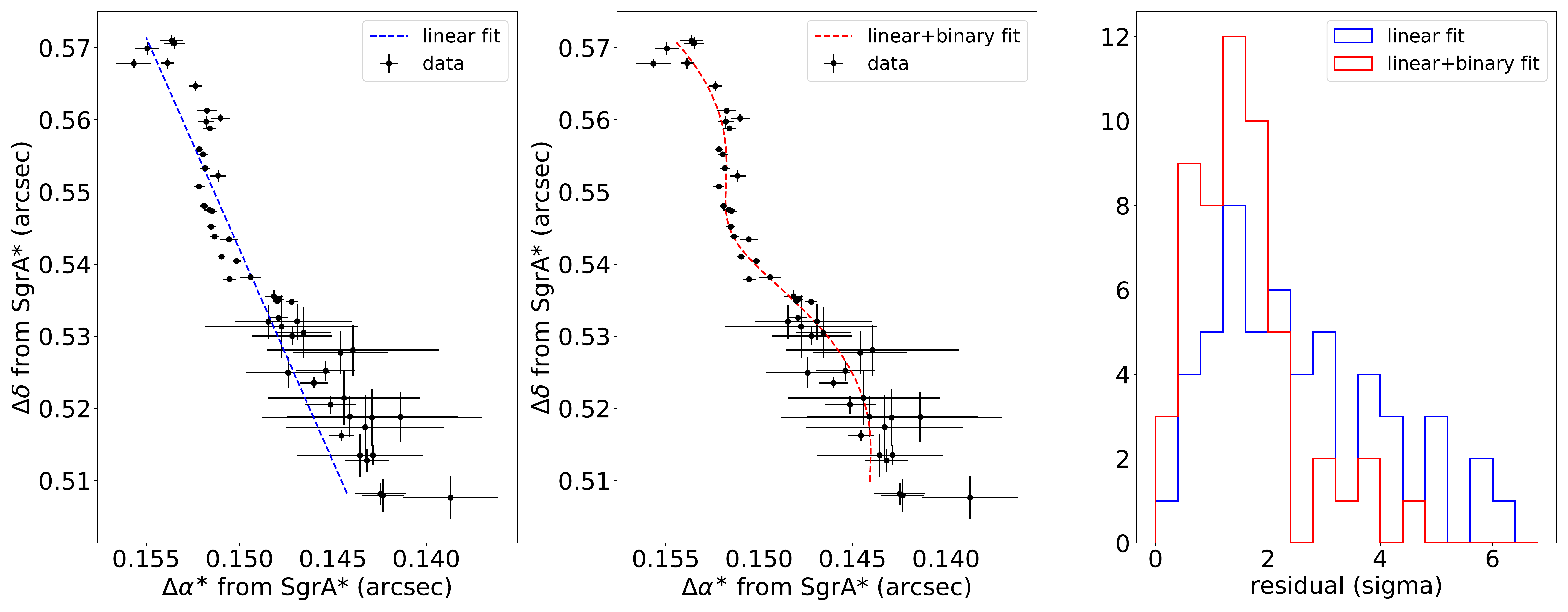}
\caption{S0-27's sky position over time with two different models overlaid. 
The \emph{left} panel uses a linear motion model and the \emph{middle} panel uses a linear+binary motion model. 
The fitting residuals between the model and the observations are plotted in the \emph{right} panel in unit of sigma.}
The periodic feature on the proper motion makes S0-27 a potential binary candidate. 

\label{fig:binary}
\end{figure*}

In \S \ref{sec:acc}, we identified 76 non-SMBH-acceleration  sources with significant acceleration inconsistent with the gravitational force of only the SMBH.
These non-SMBH-acceleration sources could be explained by  binarity, microlensing, or unrecognized confusion.
Here we use S0-27 as an example to explore the potential for binarity. 
We use two models to fit its proper motion.

Model A includes only linear motion with 4 free parameters:
\begin{list}{\labelitemi}{\leftmargin=1em}
\item $\mu_{\alpha^{\ast}}$: The proper motion of the star in X direction
\item $\mu_{\delta}$: The proper motion of the star in Y direction
\item $\alpha^{\ast}_0$: The fiducial position of the star in X direction
\item $\delta_0$: The fiducial position of the star in Y direction.
\end{list}
Then, the astrometric positions $\alpha^{\ast}$ and $\delta$ are given by:
\begin{equation}
\begin{split}
\alpha^{\ast}(t) & = \alpha^{\ast}_0 + \mu_{\alpha^{\ast}} (t-t_0)\\
\delta(t) & = \delta_0 + \mu_{\delta} (t-t_0)\\
\end{split}
\end{equation}
Here $t_0$ is fiducial time, chosen to be 1990. 

Model B includes linear motion plus binarity, which adds 6  parameters, with 11 free parameters in total: $\mu_{\alpha^{\ast}}$, $\mu_{\delta}$, $\alpha^{\ast}_0$, $\delta_0$, $\omega$, $\Omega$, $i$, $e$, $t_p$, $P$, $a$. Here we follow the models in \cite{Koren_2016}, which is summarized as follows:

\begin{list}{\labelitemi}{\leftmargin=1em}
\item $\mu_{\alpha^{\ast}}$: The proper motion of the system in X direction
\item $\mu_{\delta}$: The proper motion of the system in Y direction
\item $\alpha^{\ast}_0$: The fiducial position of the system in X direction
\item $\delta_0$: The fiducial position of the system in Y direction
\item $\omega$: The argument of periastron of the primary star's orbit in degrees.
\item $\Omega$: The longitude of the ascending node of the secondary star's orbit in degrees.
\item $i$: The inclination of the system in degrees.
\item $e$: The eccentricity of the Keplerian in orbit.
\item $t_p$: The time of periastron passage in year.
\item $P$: The period of the Keplerian orbit in year.
\item $a$: The photometric semi-major axis in mas. 
\end{list}

Then astrometric positions $x$ and $y$ are measured by:
\begin{equation}
\begin{split}
\alpha^{\ast}(t) & = \alpha^{\ast}_0 + \mu_{\alpha^{\ast}} (t-t_0) + BX(t) + GY(t)\\
\delta(t) & = \delta_0 + \mu_{\delta} (t-t_0) + AX(t) + FY(t)\\
\end{split}
\end{equation}
where A, B, F, and G are the Thiele-Innes constants \cite{vandeKamp_1967}; X(t) and Y(t) are the elliptical rectangular coordinates.

We use PyMultiNest \citep{Buchner_2014, Feroz_2008, Feroz_2009} to explore the parameter space  as it efficiently handles degeneracies inherent to binary orbit fitting.
The priors and posteriors for the two models are presented in Table \ref{tab:linfit} and Table \ref{tab:bifit}.

The best-fit models for S0-27 are shown in Figure \ref{fig:binary}, where the left panel is from the model A and the middle panel is from model B. 
Model B, with linear motion plus binarity, is a much better fit as it reduces the Bayesian information criterion (BIC) from 944 to 395, and increases the log of the likelihood from -463 to -172.

From the linear+binary model, we are able to constrain the semi-major axis $a$ and the period $P$.
If we assume the primary component is a black hole, the photometric semi-major axis $a$ would equal to the secondary component's semi-major axis. 
If we further assume the primary component is much more massive than the secondary component, then we can use $a$ as the total semi-major axis of the binary system.
Using Kepler's law, the total mass of this binary system can be calculated using the following equation: 

\begin{equation}
\frac{M}{[M_{\odot}]} = 
\left( \frac{a}{[mas]} \frac{d}{[kpc]}  \frac{1}{| \cos i |} \right)^3
\left( \frac{[yr]}{P} \right)^2
\end{equation}

With the best fit solutions in Table \ref{tab:bifit}, the total mass of the system is $\sim$ 265 $\pm$ 40 M$_{\odot}$. 
This is significantly larger than what we expect for a binary system.
However we made several simplistic assumptions that, if broken, would lower the mass.
In the future, improved constrains on the total system mass will be derived by combining our astrometric measurements with multi-epoch radial velocity measurements. 
As astrometric monitoring of the Galactic Center continues, more candidate astrometric binaries are likely to be detected and we will be able to constrain the binary fraction at the Galactic Center.

\section{Summary}
\label{summary}

The Galactic Center astrometric precision and accuracy has been increased by improving the cross-epoch alignment of starlists with the following major changes:
(1) A magnitude dependent additive error $\sigma_{add}$ is implemented for all AO epochs to create a 
    standard $\chi^2$ distribution.  
(2) A higher order local distortion map is made for 8 non-\textit{06-14} alignment data sets 
    in 2005, 2015, 2016 and 2017. 
(3) Potential confusion events are removed based on stars' proper motion.
(4) Artifact edge sources coming from elongated PSF wings are excluded from our final sample.
(5) We use jackknife to derive robust proper motion uncertainties.

These new astrometric methods produce both more precise and more accurate stellar proper motions ( $\sigma_{V}$ reduced by 40\%) as compared with our previous work \citep[e.g.][]{Boehle_2016}. 
Among the final sample of 1184 stars, we have identified 24 significantly accelerating sources with 3.5 potential contaminants, among them 15 are reported for the first time. 
We have constructed a much more stable reference frame - the position and velocity of Sgr A* derived from S0-2's orbit is both more precise and more accurate, by more than a factor of 2.

This improved astrometry will help answer many open questions in the GC. 
For example, with a better measurement of proper motion, especially significant acceleration, the young stars can be classified as disk and off-disk stars more easily, which will help with understanding the star formation history in the GC. 
Tests of General Relativity with S0-2 will be significantly improved. 
With longer time baseline, we will even be able to find potential binaries and microlensing candidates based on their astrometric measurements. 

\section{Acknowledgements}

We thank the staff of the Keck Observatory, especially Randy Campbell, Jason Chin, Scott Dahm, Heather Hershey, Carolyn Jordan, Marc Kassis, Jim Lyke, Gary Puniwai, Julie Renaud-Kim, Luca Rizzi, Terry Stickel, Hien Tran, Peter Wizinowich, Carlos Alvarez, Greg Doppman and current and former directors, Hilton Lewis and Taft Armandroff for all their help in obtaining observations.
We acknowledge support from the W. M. Keck Foundation, the Heising Simons Foundation, and the National Science Foundation (AST-1412615, AST-1518273).
The research leading to these results has received funding from the European Research Council under the European Union's Seventh Framework Programme (FP7/2007-2013) / ERC grant agreement n$^{\circ}$ [614922].
The W.M. Keck Observatory is operated as a scientific partnership among the California Institute of Technology, the University of California and the National Aeronautics and Space Administration.  
The Observatory was made possible by the generous financial support of the W. M. Keck Foundation.  
The authors wish to recognize and acknowledge the very significant cultural role and reverence that the summit of Maunakea has always had within the indigenous Hawaiian community.  
We are most fortunate to have the opportunity to conduct observations from this mountain.

\facilities{Keck Observatory}

\software{
AstroPy \citep{astropy_2013},
Matplotlib \citep{matplotlib_2007},
SciPy \citep{scipy_2001}
}

\appendix
\subsection{Central Arcsecond Astrometry Correction} 
\label{sub:central_arcsecond_astrometry_correction}

We refine our method for local distortion correction 
in the central arcsecond in order to include many additional stars with motions on the sky that exceed a linear model. This refined correction is applied only to the observed positions of S0-2 (Do et al. in prep.). The correction procedure incorporates orbital fits to seven stars close to Sgr A* as well as linear and acceleration model fits to all other stars in the central arcsecond  with $r < 1''$ across all epochs of observations (1995 -- 2018).

We fit linear, acceleration, and orbital models to the stars located in the central arcsecond. We first generated orbital models for seven stars (S0-1, S0-3, S0-5, S0-16, S0-19, S0-20, S0-38). These seven stars are those for which we have radial velocity measurements and whose orbital motions include a turning point in astrometry. S0-2, the star of interest, is excluded from the sample of orbital stars used to calculate the correction. For the remaining stars within the central arcsecond around Sgr A* ($r < 1''$), we fit both linear and accelerating models to their measured astrometric positions. We calculated the $\chi^2_{\text{red}}$ statistic for the linear and acceleration fits of each of these stars. We selected a linear model for those stars that had a lower $\chi^2_{\text{red}}$ under the linear model compared to the acceleration model. An acceleration model was used otherwise. We next removed those stars that had $\chi^2_{\text{red}} > 10$ in all cases since their motions were not well fit by either the linear or acceleration models. We also removed stars detected in fewer than 28 epochs since these stars may not have enough astrometric detections for well-constrained fits. Ultimately, we obtained one star in the central arcsecond with astrometric measurements well fit by a linear model and 19 stars with astrometric measurements well fit by an acceleration model. Along with the seven orbital model stars, we obtained a total of 27 stars for calculating the central arcsecond astrometry correction.

We next calculated the correction to S0-2's astrometric measurement using the stars detected in the central arcsecond in each epoch. For each star in the central arcsecond astrometry correction sample, we calculated the residual of each measured astrometric positions from the star's respective proper motion model (i.e.: $\text{measured position} - \text{model position}$). Uncertainties on the astrometry residual included uncertainty in the astrometric position (including both positional and alignment errors) and uncertainty in each star's respective proper motion model. For each epoch, we then calculated the weighted mean of the astrometry residuals of all central arcsecond astrometry correction stars detected in that epoch, $\bar{x}, \bar{y}$, defined in the following way:
\begin{eqnarray*}
    \bar{x} = \frac{\sum{w_{x,i} x_i}}{\sum{w_{x,i}}}
\end{eqnarray*}
Here, $w_{x,i}$ represents the weight on each star's measured position $x_i$. The weights, $w_{x,i}$ and $w_{y,i}$, were calculated from the uncertainty on the astrometric differences, $\sigma_{x,i}$ and $\sigma_{y,i}$: $w_{x,i} = 1 / \sigma_{x,i}^2$. In each epoch, the weighted mean of the residuals was subtracted from S0-2's astrometry measurements to derive the distortion corrected position for S0-2.

Uncertainties on the astrometric residuals were calculated by bootstraps. In each epoch, we constructed 1000 bootstrap trials. Each bootstrap trial had a full sample, drawn randomly with replacement, for that epoch's stellar astrometry differences and associated uncertainties. The weighted mean was then calculated on each bootstrap trial. The uncertainty on the weighted mean astrometric difference for each epoch was next calculated as half of the median-centered $1 \sigma$ (i.e. 68\%) range of the bootstrap trial means for the epoch.

\begin{figure*}[htp]
  \plotone{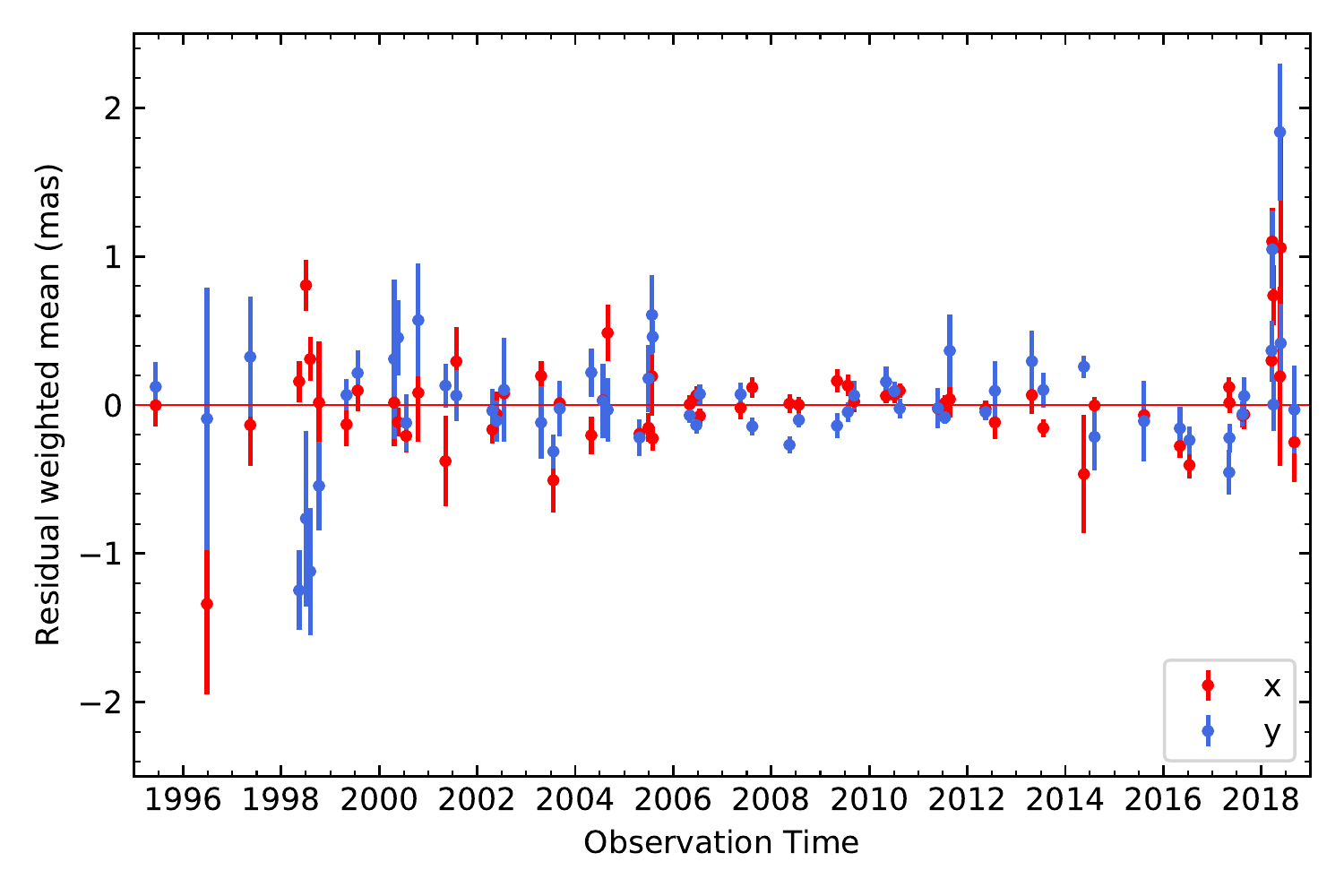}
  \caption{Residuals in $x$ and $y$ derived for S0-2's position, using 27 stars in the central arcsecond fitted to linear, accelerating, or orbital motion models. These residuals were subtracted from S0-2's astrometric measurements to derive the distortion corrected positions for S0-2.}
  \label{fig:cent_arcsec_resids}
\end{figure*}

The astrometric residuals used to correct the observed astrometric measurements of S0-2 when deriving its orbit are shown in Figure~\ref{fig:cent_arcsec_resids}. In speckle holography epochs (1995 -- 2005), median absolute corrections are 0.162 mas and 0.173 mas in $x$ and $y$, respectively, with median bootstrap uncertainties of 0.153 mas and 0.228 mas in $x$ and $y$, respectively. Median alignment uncertainties for S0-2 before local distortion correction in speckle holography epochs is 0.93 mas and 0.76 mas in $x$ and $y$, respectively. In 2005 AO epochs, median absolute corrections are 0.189 mas and 0.319 mas in $x$ and $y$, respectively, with median bootstrap uncertainties of 0.089 mas and 0.167 mas in $x$ and $y$, respectively. Median alignment uncertainties for S0-2 before local distortion correction in 2005 AO epochs is 0.21 mas and 0.32 mas in $x$ and $y$, respectively. In 2006--2014 AO epochs (i.e. \textit{06--14} setup), median absolute corrections are 0.063 mas and 0.098 mas in $x$ and $y$, respectively, with median bootstrap uncertainties of 0.063 mas and 0.072 mas in $x$ and $y$, respectively. Median alignment uncertainties for S0-2 before local distortion correction in 2006--2014 AO epochs is 0.14 mas and 0.15 mas in $x$ and $y$, respectively. In 2015--2018 AO epochs, median absolute corrections are 0.251 mas and 0.222 mas in $x$ and $y$, respectively, with median bootstrap uncertainties of 0.099 mas and 0.177 mas in $x$ and $y$, respectively. Median alignment uncertainties for S0-2 before local distortion correction in 2015--2018 AO epochs is 0.26 mas and 0.26 mas in $x$ and $y$, respectivel.



\bibliography{gc}
\end{document}